\documentclass{article}
\usepackage{amssymb,amsfonts,amsmath,stmaryrd}
\usepackage{cite,enumerate,float,indentfirst}
\usepackage{color}
\usepackage{comment}
\usepackage{url}

\def\be{\begin{eqnarray}}
\def\ee{\end{eqnarray}}
\def\nn{\nonumber}

\def\MD{\hbox{Md}}
\def\dMD{{^\vee\hbox{Md}}}
\def\qDv{{^\vee\!\hbox{qD}}}
\def\qD{\hbox{qD}}

\def\MD{\hbox{Md}}
\def\dMD{{^\vee\hbox{Md}}}
\def\MDv{{^\vee\hbox{Md}}}
\def\qDv{{^\vee\!\hbox{qD}}}
\def\qD{\hbox{qD}}
\def\Adj{\text{Adj}}

\usepackage{ dsfont } 

\def\be{\begin{eqnarray}}
\def\ee{\end{eqnarray}}
\def\nn{\nonumber}

\def\p{\partial}

\newcommand{\beq}{\begin{equation}}
\newcommand{\eeq}{\end{equation}}
\newcommand{\beqa}{\begin{eqnarray}}
\newcommand{\eeqa}{\end{eqnarray}}

\def\ve{\varepsilon}
\def\ep{\varepsilon}
\def\a{\alpha}

\hoffset= - 1.0in         

\begin{document}

\title{\vspace{1cm}{\Large {\bf Vogel's universality and Macdonald dimensions}\vspace{.2cm}}
	\author{\ {\bf Liudmila Bishler\footnote{bishlerlv@lebedev.ru}}
	\date{ }
}}

\maketitle

\vspace{-4.5cm}

\begin{center}
	\hfill FIAN/TH-16/25
\end{center}

\vspace{3.0cm}

\begin{center}
	{\small {\it Lebedev Physical Institute, Moscow 119991, Russia}}\\
\end{center}

\vspace{.1cm}

\begin{abstract}
We discuss algebraic universality in the sense of P.~Vogel for the simplest refined quantity, the Macdonald dimensions. The main known source of universal quantities is given by Chern-Simons theory.
Refinement of Chern-Simons theory means introducing additional parameters. At the level of symmetric functions, the refinement is the transition from the Schur functions to the Macdonald polynomials. We consider the Macdonald polynomials associated with the simple Lie algebras, define Macdonald dimensions and dual Macdonald dimensions, and present a universal formula for them that unifies these quantities for algebras associated with simply laced root systems. We also consider mixed Macdonald dimensions that depend on two different root systems.
\end{abstract}

\section{Introduction}

For more than two decades, P.\,Vogel \cite{Vogel95,Vogel99} worked on the construction of a Universal Lie algebra which would incorporate all simple Lie algebras. This idea has not found its ultimate realization as an actual algebra; however, a concept of \textbf{universal quantity} has been created. We call a quantity universal if it can be described as a rational function of Vogel parameters. Known universal quantities are in some way connected with Chern-Simons theory, which means that various quantities in Chern-Simons theory with a gauge group $G$ associated with structures of a simple Lie algebra $g$ have a universal description. One can hypothesize that universality belongs to gauge theories and quantities within them, rather than algebraic structures of Lie algebras.  

Known universal quantities are associated only with the adjoint representation and its descendants. This is quite natural, since the structures of representations of distinct simple Lie algebras are too different. Typical universal quantities are: the Chern-Simons partition function \cite{MkrtVes12,Mkrt13,KreflMkrt,M2}, the dimension \cite{Vogel99} and quantum dimension \cite{Westbury03,MkrtQDims} of the adjoint representation, eigenvalues of the second and higher Casimir operators \cite{LandMan06,MkrtSergVes,ManeIsaevKrivMkrt, IsaevProv, IsaevKriv,IsaevKrivProv} in these representations, the volume of simple Lie groups \cite{KhM},
the HOMFLY-PT knot/link polynomial colored with adjoint representation \cite{MMM,MM,BM4strand} and the Racah matrix involving the adjoint representation and its descendants \cite{MM,ManeIsaevKrivMkrt, IsaevProv, IsaevKriv,IsaevKrivProv,KLS}. 

Description of universal quantities is based on three Vogel's parameters: $\mathfrak{a}$, $\mathfrak{b}$ and $\mathfrak{c}$, such that one can scale all of them at once with an arbitrary constant. One usually chooses one of the parameters, $\mathfrak{a}$ to be -2. Note that these parameters are usually denoted as $\alpha$, $\beta$ and $\gamma$, however, we denote them differently, since we use $\alpha$ to denote roots of a root system.
Vogel's parameters for simple Lie algebras are listed in Table \ref{vogelparm}.
\begin{table}[H]
\centering
\begin{tabular}{|c|c|c|c|c|c|}
\hline
Root system & Lie algebra & $\mathfrak{a}$ & $\mathfrak{b}$ & $\mathfrak{c}$ & $\mathfrak{t} = \mathfrak{a}+\mathfrak{b}+\mathfrak{c}$ \\
\hline
$A_n$ & ${sl}_{n+1}$ & $-2$ & $2$ & $n+1$ & $n+1$ \\
$B_n$ & ${so}_{2n+1}$ & $-2$ & $4$ & $2n-3$ & $2n-1$ \\
$C_n$ & ${sp}_{2n}$ & $-2$ & $1$ & $n+2$ & $n+1$ \\
$D_n$ & ${so}_{2n}$ & $-2$ & $4$ & $2n-4$ & $2n-2$ \\
$G_2$ & ${g}_2$ & $-2$ & $\frac{10}{3}$ & $\frac{8}{3}$ & $4$ \\
$F_4$ & ${f}_4$ & $-2$ & $5$ & $6$ & $9$ \\
$E_6$ & ${e}_6$ & $-2$ & $6$ & $8$ & $12$ \\
$E_7$ & ${e}_7$ & $-2$ & $8$ & $12$ & $18$ \\
$E_8$ & ${e}_8$ & $-2$ & $12$ & $20$ & $30$ \\
\hline
\end{tabular}
\caption{Vogel's parameters}
\label{vogelparm}
\end{table}

There were attempts to extend the notion of universality to the refined Chern-Simons theory \cite{AgSh1,AgSh2}. In practice, it was established in \cite{KS} for the classical Lie algebras and in \cite{AM1,Mane} for the exceptional algebras that the partition function of the refined Chern-Simons theory is universal only for the simply laced algebras. Note that there is another universal formula, which unifies knot hyperpolynomials for the root systems $A_1$, $A_2$, $D_4$, $E_6$, $E_7$, $E_8$ \cite{ChE}, though it could have also included  $G_2$ and $F_4$ \cite{DG}, but does not, because these two root systems are not simply laced.

Further study of the refinement confirmed that algebraic universality manifests itself specifically in the quantities from Chern–Simons theory, which are in fact knot invariants. The analysis has been extended to the case of Macdonald dimensions \cite{BM}, which we discuss in detail in the present article, and hyperpolynomial of Hopf link \cite{BMM}. However in refined Chern-Simons theory universality is preserved only for simply laced algebras.

\begin{table}[h]
    \centering
    \begin{tabular}{|c||c|}
        \hline
        &\cr
        Quantum dimensions & Dual quantum dimensions \cr
        &\cr
        \hline
        &\cr
        $\displaystyle{ \qD_{\lambda}^{R}  : = \chi_{\lambda}^R\left(x  = q^{2\rho} \right) =\prod_{\alpha\in R_+} \frac{[(\alpha,\lambda+\rho)]_q}{[(\alpha,\rho)]_q}}$ & $\displaystyle{  \qDv_{\lambda}^{R}: = \chi_{\lambda}^R\left(x  = q^{2r} \right)=
        \prod_{\alpha \in R_+}\prod_{j=1}^{(\alpha^{\vee},\lambda)} \frac{[(\rho,\alpha^{\vee})+j]_q}{[(\rho,\alpha^{\vee})+j-1]_q}}$ \cr
        &\cr
        \hline
        \hline
        &\cr
        Macdonald dimensions & Dual Macdonald dimensions \cr
        &\cr
        \hline
        &\cr
        $\displaystyle{\MD^R_\lambda:= P^{R}_{\lambda}\left(x = q^{2\rho_k}\,|\,t_\alpha^2\,|\,q^2,t^2\right)}$ &
        $\displaystyle{\dMD_{\lambda}^{R} =  P^{R}_{\lambda}\left(x = q^{2r_k}\,|\, t_{\alpha}^2\,|\, q^2, t^2 \right)=\prod_{\alpha\in R_+} \, \prod_{j=1}^{(\alpha^{\vee},\lambda)}\,  \frac{\left\{t_{\alpha} q^{(\rho_k,\alpha^{\vee})+j-1}\right\}}{\left\{q^{(\rho_k,\alpha^{\vee})+j-1}\right\}}}$ \cr
        &\cr
        \hline
    \end{tabular}
    \caption{Quantum and Macdonald dimensions and their counterparts}
    \label{alldimtable}
\end{table}

In Chern-Simons theory quantum dimensions are Wilson loop averages of unknotted loop. In representation theory they are characters of irreducible representations taken at the point of Weyl vector $\rho$. The nice thing about quantum dimensions is that they factorize and admit universalization in adjoint representation \cite{Westbury03,MkrtQDims} and its descendants. Characters of irreducible representations also factorize at the dual Weyl vector.

The counterpart of quantum dimensions in refined Chern-Simons theory are Macdonald dimensions --- Macdonald polynomials at the point of refined Weyl vector. However, they do not factorize. Only dual Macdonald dimensions admit factorization. Dual Macdonald dimensions differ from Macdonald dimensions for non simply laced root systems, which means that Macdonald dimensions for $A_n$, $D_n$ and $E_6$, $E_7$, $E_8$ root systems do factorize and as it turns out admit universalization. We present the universal formula for Macdonald dimensions in adjoint representation of simply laced root systems (see eq. (\ref{main1}). We summarize the key elements discussed in this paper in Table 2.

This paper is an extended version of \cite{BM}. We go into more details of the definitions of Macdonald polynomials and of Macdonald dimensions. We also consider mixed Macdonald dimensions that depend on two different root systems.

The structure of the paper is as follows. In section \ref{MacSec}
we start with the introduction of Macdonald polynomials and of all the components of their definition: admissible pairs, scalar product, dominance order, parameters and basis. We want to emphasize the difference between the Macdonald polynomials for the root system $A_n$ and for other classical root systems. In section \ref{SchursSection}, we discuss the Schur functions: a natural basis of Macdonald polynomials. In section \ref{DimsSec}, we define quantum and Macdonald dimensions and their dual versions, discuss factorization of these quantities and list them in adjoint representations of simple Lie algebras and admissible pairs $(R,R)$. In section \ref{MixedMacSec}, we consider the Macdonald dimensions depending on two different root systems. Finally in section \ref{UniSec}, we discuss universality in Chern-Simons theory, in particular, universal formulas for the quantum and Macdonald dimensions and possibilities for further universalization.

\paragraph{Notation and comments.}
\begin{itemize}
\item We use the letter $\alpha$ to denote the roots of the root system $R$, $R_+$ is a set of  positive roots.
\item
In variance with the original work by I. Macdonald \cite{Mac} and subsequent papers on the subject \cite{MacConj,CherednikConj,CherednikDAHA,Koorn},
we use symmetric quantum numbers, which allows us to present the results in a shorter and more elegant form: in our notation, the Macdonald polynomials depend on the squares of the original Macdonald parameters:
\begin{equation}
    q \,\rightarrow \, q^2, \,\, t \,\rightarrow \, t^2,\,\,  t_{\alpha}^2 \,\rightarrow \, t_{\alpha}^2
\end{equation}
and the brackets and quantum numbers are defined to be
\begin{equation}
    \{x\} = x-x^{-1}, \quad \{x\}_{_+}=x+x^{-1}, \quad[n]_q = \frac{q^n-q^{-n}}{q-q^{-1}}, \quad [n]_t = \frac{t^n-t^{-n}}{t-t^{-1}}.
\end{equation}

\item When we say that some symmetric polynomial that depends on variables $x_1, \dots, x_n$ is taken at the point $q^{\rho}$, where $\rho = (\rho_1, \dots,\rho_n)$ is the Weyl vector, we mean that one should make the substitution in this symmetric polynomial
\begin{equation}
    x_i = q^{\rho_i}.
\end{equation}
The number of variables $x_i$ and the length of the Weyl vector coincide and are equal to the dimension of the Euclidean space where the root lattice is embedded.

\item We use $\mathfrak{a}$, $\mathfrak{b}$ and $\mathfrak{c}$ as well as $\mathfrak{t} = \mathfrak{q}+\mathfrak{b}+\mathfrak{c}$ to denote Vogel parameters.

\item Vogel parameters (Table \ref{vogelparm}) are based on the minimal normalization of the roots, which means that the length of the longest root $(\alpha_l,\alpha_l) = 2$.

\end{itemize}

\section{Macdonald polynomials \label{MacSec}}
Macdonald polynomials $P_{\lambda}^{(R,S)} (x\,|\,t_{\alpha}\,|\,q,t)$ associated with a pair of root systems $(R,S)$ were defined by I.\,Macdonald \cite{Mac}, and are a generalization of famous Macdonald polynomials $M_{\lambda}(x\,|\,q,t)$ \cite{Mac0,Mac01}, which turn out to be associated with the root system $A_n$:
\begin{equation}
    P_{\lambda}^{(A_n,A_n)} (x \,|\,q,t) = M_{\lambda}(x\,|\,q,t).
\end{equation}
Polynomials $P_{\lambda}^{(R,S)} (x\,|\,t_{\alpha}\,|\,q,t)$ are symmetric under action of the Weyl group $W = W_R = W_S$. They are enumerated with dominant weights $\lambda$ and depend on parameters $q$ and $t$, and additionally on $t_{\alpha}$ that are associated with roots of distinct lengths.

\subsection{Root systems \label{RootSystemsSection} }
It is convenient to use root systems to  classify simple Lie algebras. All simple Lie algebras are associated with one of the following root systems:
\begin{equation}
    A_n, \ \  B_n, \ \  C_n, \ \  D_n, \ \  E_6, \ \ E_7, \ \ E_8, \ \ F_4,\ \ G_2.
\end{equation}
Root systems $ A_n, $ $  B_n,$ $  C_n, $ $  D_n$ are called classical and they are associated with algebras $sl_{n+1}$, $so_{2n+1}$, $sp_{2n}$ and $so_{2n}$ correspondingly. Root systems $E_6, $ $ E_7,$ $ E_8, $ $ F_4,$ $G_2$ are called exceptional and are associated with algebras $e_6$,  $e_7$, $e_8$, $f_4$ and $g_2$.

\vspace{10pt}
To define a root system one starts with the Euclidean space $V$ with the \textbf{orthogonal basis} $\ve$:
\begin{equation}
    (\ve_i,\ve_j) = \delta_{ij}.
    \label{Vspace}
\end{equation}
One can define a reflection on $V$:
\begin{equation}
    s_{a}(v) := v- 2\frac{(v,a)}{(a,a)}a = v - (v,a)a^{\vee},
\end{equation}
where $v\in V$, $a\in V \, \backslash \, \{0\}$, $\a^{\vee} = 2a/(a,a)$.

We call $R \subset V \, \backslash \, \{0\} $ a root system if for any $\alpha,\, \beta \in R$
\begin{align}
   R \text{ spans } V, \quad\quad
    s_{\alpha} (\beta) \in R, \quad\quad
    (\alpha^{\vee}, \beta) \in \mathbb{Z},
\end{align}
$\alpha$, $\beta$ are roots of a root system $R$, $\alpha^{\vee}$ are coroots $\alpha^{\vee} = 2\alpha/(\alpha,\alpha)$.

Weyl group $W_R$ of root system $R$ is the subgroup of the orthogonal group $O(V)$ generated by reflections from roots $\alpha \in R$.

In addition to the orthogonal basis there are two other natural bases on the space of a root system: basis of simple roots and basis of fundamental weights.

\textbf{Simple roots} $\{ \alpha_i \, |\, i\in I\} \subset R$ are a basis of $R$ if $\forall \alpha \in R$ we can expand $\alpha = c_i\,\alpha_i$ so that all $c_i$ are of the same sign or equal to zero. Other roots can be expressed as
\begin{equation}
    \alpha = \sum_{i\in I} c_i\,\alpha_i.
\end{equation}
\noindent
Positive roots  $\alpha \in R_+ $ are the roots with non-negative coefficients $c_i$:
\begin{equation}
    R_{+} = \{ \alpha \in R \,|\, c_i \geq 0 \}.
\end{equation}

\textbf{Fundamental weights} $\omega_i$ are defined as
\begin{equation}
    (\omega_i,\alpha_j^{\vee}) = \delta_{ij}.
\end{equation}
The combinations of fundamental weights $\lambda$ enumerate irreducible representations of Lie algebras as well as Macdonald polynomials:
\begin{equation}
    \lambda = \sum \lambda_i \omega_i, \quad \lambda_i \in \mathbb{N}_0.
\end{equation}
They are called dominant weights.

\subsubsection{Root system $A_n$ (algebra $sl_{n+1}$)}

Roots, positive roots and Weyl group $W$ of the root system $A_n$ are the following:
\begin{equation}
    R = \left\{\ve_i-\ve_j\,|\, 1\leq i \neq j \leq n+1\right\}, \quad\quad W \simeq S_{n+1} \end{equation}
 \begin{equation}
    R^{+} = \left\{\ve_i-\ve_j\,|\, 1\leq i < j \leq n+1\right\}. \label{rootsAplus}
    \end{equation}

The Schur-Weyl duality establishes correspondence between representations of $sl_N$ and the permutation group $S_N$.
Irreducible representations of $S_N$ are enumerated with partitions $Q$ --- sequences of non-negative integers in non-increasing order:
\begin{equation}
Q = [Q_1, Q_2, \dots, Q_N], \quad Q_1\geq Q_2\geq \dots\geq Q_N, \quad Q_i \in \mathbb{N}_0,\quad |Q| = \sum_{i=1}^{N}Q_i,
\end{equation}
$l(Q)$ --- length of partition --- number of non-zero $Q_i$ in $Q$.

Symmetric functions such as power sum $p_{Q}$, monomial $m_Q$ and Schur symmetric functions are enumerated with partitions:
\begin{align}
   & p_k = \sum_{i=1}^{N} x_i^{k}, \quad p_{Q} = \prod_{i=1}^{l(Q)} p_{Q_i}, \label{psum} \\
   & m_{Q} = \sum_{\sigma\in S_N} x_{\sigma(1)}^{Q_1}x_{\sigma(2)}^{Q_2} \dots x_{\sigma(l(Q))}^{Q_{l(Q)}}, \label{monom}
\end{align}
and we are going to discuss Schur symmetric functions in detail in section \ref{SchursSection}.

Irreducible representations of $sl_{n+1}$ are also labeled with partitions. Partitions can be represented as Young diagrams and are sometimes identified with them. They are connected with dominant weights $\lambda$ in the following way:
\begin{equation}
    \begin{aligned}
      &  \lambda = \sum_{i=1}^{n} \lambda_i \omega_i , \\
      &  Q_{\lambda} = [Q_1, Q_2, \dots, Q_{n+1}], \quad \\
      &  Q_i = \sum_{j=i}^{n} \lambda_j.
    \end{aligned}
    \label{weigthA}
\end{equation}

The fundamental weight $\omega_i$ in $(n+1)$-dimensional orthogonal basis $\ep_i$ is
\begin{equation}
    \omega_i = \frac{n+1-i}{n+1} (\ep_1+\dots+\ep_i)-\frac{i}{n+1}(\ep_{i+1}+\dots+\ep_{n+1}).
\end{equation}
\noindent
To enumerate irreducible representations of algebras $so_N$ and $sp_N$ one can use the highest weights of representations in the orthogonal basis.

\subsubsection{Root system $B_n$ (algebra $so_{2n+1}$)}

Roots, positive roots and Weyl group $W$ of the root system $B_n$ are the following:
\begin{equation}
    R = \left\{\pm \ve_i | 1\leq i\leq n \right\} \cup \left\{\pm \ve_i\pm \ve_j| 1\leq i < j \leq n\right\}, \quad\quad W \simeq S_{n} \ltimes (\{\pm 1\})^n\end{equation}
\begin{equation}
    R^+ = \left\{\ve_i | 1\leq i\leq n \right\} \cup \left\{ \ve_i\pm \ve_j| 1\leq i < j \leq n\right\}. \label{rootsBplus}
    \end{equation}

Fundamental weights in the orthogonal $n$-dimensional basis $\ep_i$ are
\begin{align}
  &  \omega_i = \sum_{k=1}^{i} \ep_k \quad \text{for} \quad i=1,\dots,n-1, \\
  & \omega_n = \frac{1}{2} \sum_{k=1}^{n} \ep_k \label{OmegaNB}.
\end{align}
And dominant weight $\lambda = \sum \lambda_i \omega_i$ in the orthogonal basis has the following coordinates:
\begin{equation}
  \lambda^{B_n}_{\ve} =   \left [\lambda_1+\lambda_2+\dots+\frac{\lambda_n}{2},\lambda_2+\dots+\frac{\lambda_n}{2},\dots,\frac{\lambda_n}{2} \right].
   \label{weightB}
\end{equation}
In this paper we treat dominant weights $\lambda$ in the orthogonal basis as partitions, even though there are weights with half-integer coordinates. The same works for the root system $D_n$.

\subsubsection{Root system $C_n$ (algebra $sp_{2n}$)}

Roots, positive roots and Weyl group $W$ of the root system $C_n$ are the following:
\begin{equation}
    R =  \left\{\pm  \ve_i\pm \ve_j| 1\leq i < j \leq n\right\}\cup\left\{\pm 2\ve_i | 1\leq i\leq n \right\} , \quad\quad W \simeq S_{n} \ltimes (\mathbb{Z}/2\mathbb{Z})^n, \label{rootsC}
    \end{equation}
    \begin{equation}
    R_{+} =  \left\{\ve_i\pm \ve_j| 1\leq i < j \leq n\right\}\cup\left\{2\ve_i | 1\leq i\leq n \right\}.\label{rootsCplus}
    \end{equation}
We want to stress that roots (\ref{rootsC}) together with the scalar product $(\ve_i,\ve_j)=\delta_{ij}$ are \textit{not} in the minimal normalization. We use them in this paper in order to be consistent with other works on Macdonald polynomials \cite{Koorn}.

Fundamental weights in $n$-dimensional orthogonal basis $\ep_i$ are
\begin{equation}
    \omega_i = \sum_{k=1}^i \ep_k \quad \text{for}\quad i=1,\dots,n
\end{equation}
and general weight $\lambda = \sum \lambda_i \omega_i$ in the orthogonal basis is the integer partition:
\begin{equation}
  \lambda^{C_n}_{\ve} =   \left [\lambda_1+\lambda_2+\dots+\lambda_n,\lambda_2+\dots+\lambda_n,\dots,\lambda_n \right].
   \label{weightC}
\end{equation}

\subsubsection{Root system $D_n$ (algebra $so_{2n}$)}
Roots, positive roots and Weyl group $W$ of the root system $D_n$ are the following:
\begin{equation}
    R = \left\{\pm \ve_i\pm \ve_j| 1\leq i < j \leq n\right\}, \quad\quad W \simeq S_{n} \ltimes (\mathbb{Z}/2\mathbb{Z})^{n-1},\end{equation}
    \begin{equation}
    R^+ = \left\{ \ve_i\pm \ve_j| 1\leq i < j \leq n\right\}. \label{rootsDplus}
    \end{equation}
Weights in the $n$-dimensional orthogonal basis $\ep_i$ are the following:
\begin{align}
   & \omega_i = \sum_{k=1}^i \ep_k \quad \text{for} \quad i=1,\dots,n-2, \\
   &\omega_{n-1} = \frac{1}{2} \sum_{k=1}^{n-1} \ep_k -\frac{1}{2} \ep_n   ,\\
   & \omega_{n} = \frac{1}{2} \sum_{k=1}^{n} \ep_k
\end{align}
and coordinates of the dominant weight $\lambda = \sum \lambda_i \omega_i$ in the orthogonal basis $\ve_i$ are the following
\begin{equation}
  \lambda^{D_n}_{\ve} = \left [\lambda_1+\lambda_2+\dots+\frac{\lambda_{n-1}}{2}+\frac{\lambda_n}{2},\lambda_2+\dots+\frac{\lambda_{n-1}}{2}+\frac{\lambda_n}{2},\dots,\frac{\lambda_{n-1}}{2}+\frac{\lambda_n}{2}, \frac{\lambda_n}{2}-\frac{\lambda_{n-1}}{2} \right].
   \label{weightD}
\end{equation}
They can be half-integer and negative.

Roots of the exceptional roots systems are listed in the Appendix A.

\subsection{Definition of Macdonald polynomials associated with root systems}

For an admissible pair of root systems $(R,S)$   there exists a unique family of polynomials $P_{\lambda}^{(R,S)}$ which satisfy the following conditions:
\begin{align}
   & P_{\lambda}^{(R,S)} = m_{\lambda}^R +\sum_{\mu<\lambda} c_{\lambda\mu}^{(R,S)}\, m_{\mu}^R, \label{triangulardecomposition}\\
 &  \left \langle P_{\lambda}^{(R,S)},m_{\mu}^R  \right \rangle = 0 \,\,\, \text{if} \,\,\, \mu<\lambda. \label{orthogonality}
\end{align}
The resulting polynomials are mutually orthogonal:
\begin{equation}
\left  \langle P_{\lambda}^{(R,S)},P_{\mu}^{(R,S)}  \right \rangle = 0, \quad\quad \lambda \neq \mu.
\end{equation}

In this definition we need to specify the following symbols and operations.
\begin{itemize}
    \item $(R,S)$ is an \textbf{admissible pair} of root systems if they have the same Weyl group $W_R = W_S$, $R$ is irreducible, but not necessarily reduced, $S$ is irreducible and reduced. There are the following possible combinations:
\begin{align}
 &   (R,R):  \,\,\,  R = \, A_n,\,\,B_n, \,\, C_n,\,\,D_n, \,\, E_6,\,\,E_7, \,\, E_8,\,\,F_4, \,\, G_2 \\
 &   (R,R^{\vee}):  \,\,\, R = \,B_n, \,\, C_n, \,\,\, B_n^{\vee} = C_n, \,\, C_n^{\vee} = B_n \\
 & (BC_n, S): \,\,\, S= \, B_n, \,\, C_n
\end{align}
\item $\lambda$ is the \textbf{dominant weight} of the root system $R$, it enumerates polynomials $P_{\lambda}^{(R,S)}$ and is the combination of the fundamental weights $\omega_i$:
\begin{equation}
    \lambda = \sum_i \lambda_i \omega_i, \quad \lambda_i \in \mathds{N}_0,\quad (\omega_i,\alpha_j^{\vee}) = \delta_{ij}.
\end{equation}

\item $m_{\lambda}^R$ form a \textbf{basis} for polynomials $P_{\lambda}^{(R,S)}$. They are an alternative to monomial symmetric functions (\ref{monom}):
\begin{equation}
    m_{\lambda}^R = \frac{1}{|W_R^{\lambda}|} \sum_{w\in W_R} e^{w \lambda}(v), \label{MacBasis}
\end{equation}
where $W_R^{\lambda}$ is the stabilizer of $\lambda$ in Weyl group $W_R$. In the case of classical Lie algebras one can also use Schur functions (\ref{schurA})--(\ref{schurD}) as a basis for Macdonald polynomials. We discuss them in section \ref{SchursSection}.

\item The \textbf{triangular decomposition} (\ref{triangulardecomposition}) is build with the dominance order $\mu <\lambda $ on weights $\lambda$ and $\mu$:
\begin{equation}
    \lambda \geq \mu \,\, \Longleftrightarrow \,\, \lambda-\mu \in \mathbb{N} R_{+}.
\end{equation}
We discuss dominance order for classical root systems in more detail in the section \ref{DOrderSec}.
\item The \textbf{scalar product} is defined as follows
\begin{equation}
    \langle f,g \rangle = |W|^{-1} \int_T \prod_{\alpha \in R} f(\dot v) \overline{g(\dot v)} \Delta(\dot v) d \dot v,
\end{equation}
 where $\int_T d \dot v = 1$. In case of finite number of variables $x_i$, $1\leq i \leq n$ we can rewrite this formula:
\begin{equation}
     \langle f(x),g(x) \rangle = |W|^{-1} [f(x)g(x^{-1})\Delta]_0,
\end{equation}
where the operator $[\dots]_0$ means picking up the constant term of a Laurent polynomial.

The \textbf{Macdonald density} is defined as a product over all roots of the root system $R$:
 \begin{equation}
     \Delta(v) := \prod_{\alpha \in R} \frac{\left(t_{2\alpha}^{1/2}e^{\alpha}(v);q_{\alpha}\right)_{\infty}}{\left(t_{\alpha}t_{2\alpha}^{1/2}e^{\alpha}(v);q_{\alpha}\right)_{\infty}}, \quad\quad (a;q)_{\infty} = \prod_{i=0}^{\infty} (1- aq^i).
 \end{equation}

We define the weight function $\Delta$ and discuss it in detail in Appendix A. And we discuss the parameters $t_{\alpha}$ and $q_{\alpha}$ in detail in section \ref{ParmSec}.

\item One should treat the exponent in the formulas above as
\begin{equation}
    e^{\eta} (v) = e^{(\eta,v)},
\end{equation}
where $v$ is a formal vector $v = \sum v_i \ve_i$, $\ve_i$ are the elements of the orthogonal basis (\ref{Vspace}) on the Euclidean space $V$, $\eta \in V$
and variables of Macdonald polynomials $x_i$ emerge as 
\begin{equation}
    x_i = e^{v_i}.
\end{equation}

\end{itemize}

\subsection{Dominance order for classical Lie algebras \label{DOrderSec}}

In this section we are going to formulate the rules of ordering of dominant weights for classical root systems. In the case of root systems $A_n$ we are going to formulate them for partitions $Q$ (\ref{weigthA}), and for $B_n$, $C_n$ and $D_n$ for weights in orthogonal basis (\ref{weightB}), (\ref{weightC}), (\ref{weightD}). In the case of root system $C_n$ dominant weights in orthogonal basis (\ref{weightC}) are in fact partitions. In case of root systems $B_n$ and $D_n$ (\ref{weightB}), (\ref{weightD}) are not partitions, because their coordinates can be half-integer or negative. However some rules of ordering in this case repeat the ones for partitions.

\subsubsection{Dominance order for $A_n$}
Weights $\lambda^{A_n}$ and $\mu^{A_n}$ corresponding to partitions $\Lambda$ and $\text{M}$ satisfy
\begin{equation}
    \lambda^{A_n} \geq \mu^{A_n},
\end{equation}
when
\begin{align}
   & |\Lambda|\equiv|\text{M}|, \\
   &\sum_{i=1}^k (\Lambda_i -  \text{M}_i) \in \mathbb{N}_0 \quad \text{for} \quad k=1,\dots, \max(l(\Lambda),l(\text{M})). \label{BasicOrdering}
\end{align}
It means that in the case of $A_n$ all weights are broken into groups which are enumerated with partitions of integers:
\begin{equation}
    \begin{aligned}
        & |\Lambda| = 1: & & [1] \\
        & |\Lambda| = 2: & & [1,1] < [2] \\
        & |\Lambda| = 3: & & [1,1,1] <[2,1]<[3] \\
        & |\Lambda| = 4: & & [1,1,1,1] <[2,1,1]<[2,2] < [3,1] < [4] \\
        & |\Lambda| = 5: & & [1,1,1,1,1] < [2,1,1,1] < [2,2,1]<[3,1,1] < [3,2]<[4,1]<[5]
    \end{aligned}
\end{equation}
Weights associated with different $|\Lambda|$ are incomparable and do not appear in each other's Macdonald polynomials. If the condition $(\ref{BasicOrdering})$ is not applicable to two partitions, it means that we can not compare corresponding weights. For example we can't compare $[4,1,1]$ and $[3,3]$:
\begin{equation}
    m = 6: \,\,\, \dots < [3,2,1] <
    \begin{array}{c}
         [4,1,1] \\[0.2pt]
         [3,3] \\
    \end{array}
     < [4,2] < [5,1] <[6].
\end{equation}

\subsubsection{Dominance order for $B_n$}
The dominance order for two weights $ \lambda^{B_n}_{\ve} = [\Lambda_1, \Lambda_2, \dots,\Lambda_n] $ and $\mu^{B_n}_{\ve} =[\text{M}_1, \text{M}_2, \dots, \text{M}_n] $ (\ref{weightB}) in the orthogonal basis coincides with (\ref{BasicOrdering}):
\begin{equation}
    \lambda^{B_n}_{\ve}  \geq \mu^{B_n}_{\ve}   \quad \longleftrightarrow \quad \sum_{i=1}^{k} (\Lambda_i-\text{M}_i) \in \mathbb{N}_0 \quad \text{for}\quad k=1,\dots,n
\end{equation}
The dominance order divides all dominant weights into just two groups: integer partitions and half-integer partitions. One can get the second group from the first one adding last fundamental weight $\omega_n$ (\ref{OmegaNB}):
\begin{equation}
  \begin{aligned}
  & \text{integer:} 
     & & \varnothing < [1] < [1,1] <
      \begin{array}{c}
           [2] \\[0.2pt]
           [1,1,1] \\
      \end{array} < [2,1] <  \dots \\
    & \text{half-integer} =\text{integer}+\omega_n: 
     && \varnothing+\omega_n < [1]+\omega_n < [1,1] +\omega_n<
      \begin{array}{c}
           [2] +\omega_n \\[0.2pt]
           [1,1,1]+\omega_n \\
      \end{array} < [2,1]+\omega_n <  \dots
  \end{aligned}
\end{equation}
Weights from these two groups are incomparable and do not mix in Macdonald polynomials. Inside each group there is the ordering (\ref{BasicOrdering}).

\subsubsection{Dominance order for $C_n$}

The dominance order for two weights $ \lambda^{C_n}_{\ve} = [\Lambda_1, \Lambda_2, \dots,\Lambda_n] $ and $\mu^{C_n}_{\ve} =[\text{M}_1, \text{M}_2, \dots, \text{M}_n] $ (\ref{weightC}) in the orthogonal basis is the following:

\begin{align}
    \lambda^{C_n}_{\ve} \geq \mu^{C_n}_{\ve} \quad \longleftrightarrow \quad \sum \Lambda_i\,\equiv\, \sum \text{M}_i \,(\text{mod }2) \quad \text{and} \quad \sum_{i=1}^{k}(\Lambda_i - \text{M}_i) \in \mathbb{N}_0 \quad\text{for}\quad k=1,\dots,n.
\end{align}

Dominant weights are divided into two groups: partitions of even and odd integers:
\begin{equation}
    \begin{aligned}
     & \text{even integers:}
      & &  \varnothing < [1,1]<
        \begin{array}{c}
           [2] \\[0.2pt]
           [1,1,1,1] \\
      \end{array} <
      [2,1,1]<[2,2]<[3,1]<[4]< \dots\\
       & \text{odd integers:}
      & &  [1] < [1,1,1] < [2,1] < [3] < \dots
    \end{aligned}
\end{equation}

\subsubsection{Dominance order for $D_n$}
The dominance order for two weights $ \lambda^{D_n}_{\ve} = [\Lambda_1, \Lambda_2, \dots,\Lambda_n] $ and $\mu^{D_n}_{\ve} =[\text{M}_1, \text{M}_2, \dots, \text{M}_n] $ (\ref{weightD}) in the orthogonal basis is the following:
\begin{equation}
   \lambda^{D_n}_{\ve} \geq \lambda^{D_n}_{\ve} \quad \longleftrightarrow \quad \sum_{i=1}^{n-1} (\Lambda_i-\text{M}_i)\pm (\Lambda_n-\text{M}_n) \in 2\mathbb{N}_0 \quad \text{and} \quad \sum_{i=1}^{k}(\Lambda_i - \text{M}_i) \in \mathbb{N}_0\quad\text{for}\quad k=1,\dots,n.
   \label{DominanceOrderD}
\end{equation}
It means that there are four different groups of weights which are mutually incomparable:
\begin{enumerate}
\setlength{\itemsep}{-3pt} 
    \item partitions of even integers
    \item partitions of odd integers
    \item (partitions of even integers)$+\omega_n$ + (partitions of odd integers)$+\omega_{n-1}$
    \item (partitions of even integers)$+\omega_{n-1}$ + (partitions of odd integers)$+\omega_{n}$
\end{enumerate}
The first two groups are the same as for $C_n$:
\begin{equation}
    \begin{aligned}
     & \text{even integers:}
      & &  \varnothing < [1,1]<
        \begin{array}{c}
           [2] \\[0.2pt]
           [1,1,1,1] \\
      \end{array} <
      [2,1,1]<[2,2]<[3,1]<[4]< \dots\\
       & \text{odd integers:}
      & &  [1] < [1,1,1] < [2,1] < [3] < \dots
    \end{aligned}
\end{equation}
The third and the fourth groups are mixed combinations of even and odd partitions:
\begin{equation}
    \begin{aligned}
     & \text{mixed even integers:}
      & &  \varnothing+\omega_n < [1]+\omega_{n-1} < [1,1]+\omega_n <
        \begin{array}{c}
           [2]+\omega_n \\[0.2pt]
           [1,1,1]+\omega_{n-1} \\[0.2pt]
      \end{array} <
      [2,1]+\omega_{n-1}<[3]+\omega_{n-1}< \dots\\
       & \text{mixed odd integers:}
      & &  \varnothing+\omega_{n-1} < [1]+\omega_{n} < [1,1]+\omega_{n-1} <
        \begin{array}{c}
           [2]+\omega_{n-1} \\[0.2pt]
           [1,1,1]+\omega_{n} \\[0.2pt]
      \end{array} <
      [2,1]+\omega_{n}<[3]+\omega_{n}< \dots
    \end{aligned}
    \label{DominanceOrderD34}
\end{equation}
The third and fourth groups can be obtained with the ordering (\ref{BasicOrdering}) of all integer partitions, also one needs to add $\omega_n$ to partitions of even integers and $\omega_{n-1}$ to partitions of odd integers to third group and vise versa to fourth group. However the picture is in fact more complicated and one has to  check all the conditions (\ref{DominanceOrderD}) to compare partitions. For example, one can not compare partitions $[1,1,1,1]+\omega_{4}$ and $[1,1,1]+\omega_{3}$ for root system $D_4$, but $[1,1,1,1]+\omega_{n}>[1,1,1]+\omega_{n-1}$ for other $D_n$ with $n\geq 5$. Also, some partitions from (\ref{DominanceOrderD34}) can coincide for some root systems. For example $[1,1]+\omega_3 = [1,1,1]+\omega_2$ for $D_3$.

\subsection{Parameters of Macdonald polynomials \label{ParmSec}}
In contrast to  Macdonald polynomials $M_{\lambda} (x\,|\,q^2,t^2)$ \cite{Mac0,Mac01} corresponding to the admissible pair $(A_n,A_n)$ and depending on parameters $q$ and $t$, in general case we get new parameters: $q_{\alpha}$ and $t_{\alpha}$:
\begin{itemize}
    \item $q_{\alpha}$ depends on the correspondence between roots of root systems $R$ and $S$ from an admissible pair:
\begin{equation}
    q_{\alpha} = q^{ u_{\alpha}}, \quad\quad\quad u_{\alpha}:\quad \alpha/u_{\alpha} \in S, \quad \alpha \in R. \label{uparm}
\end{equation}
\item ${t_{\alpha} = q_{\alpha}^{ k_{\alpha}}}$, where $k_{\alpha}$ depends only on a length of a root $\alpha \in R$. For roots from section \ref{RootSystemsSection} and Appendix A we use the following notation:

\begin{equation}
    \begin{array}{lll}
         (\alpha,\alpha) = 1, &\quad  t_{\alpha}  = t_s, &\quad  k_{\alpha}  = k_s,\\
    (\alpha,\alpha) = 2, &  \quad  t_{\alpha}  = t, & \quad  k_{\alpha}  = k,\\
     (\alpha,\alpha) = 4, & \quad t_{\alpha}  = t_l, & \quad k_{\alpha}  = k_l,\\
     (\alpha,\alpha) = 6, & \quad t_{\alpha}  = t_3, & \quad k_{\alpha}  = k_3. \\
    \end{array}
\end{equation}
\end{itemize}

Here are all Macdonald polynomials with corresponding parameters
\begin{equation}
    \begin{array}{ll}
    (R,R):\quad  &  P_{\lambda}^{A_n} \left( x\,|\, q,t \right), \quad P_{\lambda}^{B_n} \left( x\,|\,t_s = q^{k_s} \,|\, q,t \right), \quad  P_{\lambda}^{C_n} \left( x\,|\,t_l = q^{k_l} \,|\, q,t \right), \quad  P_{\lambda}^{D_n} \left( x \,|\, q,t \right)  \\
    &\\
    & P_{\lambda}^{E_n}  \left( x\,|\, q,t \right),\quad  P_{\lambda}^{F_4} \left( x\,|\,t_s = q^{k_s} \,|\, q,t \right), \quad  P_{\lambda}^{G_2} \left( x\,|\,t_3 = q^{k_3} \,|\, q,t \right) \\
    &\\
     (R,R^{\vee}):\quad & P_{\lambda}^{(B_n,C_n)} \left( x\,|\,t_s = q^{k_s}\,|\, q,t \right), \quad P_{\lambda}^{(C_n,B_n)} \left( x\,|\,t_l = q^{k_l} \,|\, q,t \right) \\
     &\\
     (BC_n,R): \quad & P_{\lambda}^{(BC_n,B_n)} \left( x\,|\,a = q^{k_s},\, b = q^{2 k_l} \,|\, q,t \right), \quad P_{\lambda}^{(BC_n,B_n)} \left( x\,|\,a = q^{k_s/2},\, b = q^{k_l} \,|\, q,t \right) \\
    \end{array}
    \label{admissible pairs}
\end{equation}

We also want to point out that some Macdonald polynomials can be generalized with Koornwinder polynomial \cite{Koorn}. The details are in Appendix B.

\subsection{Characters of classical simple Lie algebras \label{SchursSection} }
It is natural to use characters of irreducible representations of classical Lie algebras as the basis of Macdonald polynomials  instead of (\ref{MacBasis}). In this section we want to go into some details about characters of the classical Lie algebras and Schur functions.

\subsubsection{Characters of the $sl_N$}
Characters of the $sl_N$ Lie algebra, which corresponds to the root system $A_{N-1}$ are Schur symmetric functions
\begin{equation}
    \chi_{\lambda}^{A_{N-1}} = S_{\lambda}(x_1, \dots, x_N),
\end{equation}
which can be defined with the Cauchy's bialternant formula, also called Weyl's formula:
\begin{equation}
    S_{\lambda}(x_i,\dots,x_N) = \frac{\text{det}\left(x_i^{\lambda_j+N-j}\right)}{\text{det}\left(x_i^{N-j}\right)}.
\end{equation}
In fact all dependence on $N$ in Schur functions can be hidden in variables $p_k$ (\ref{psum}), which are power sum symmetric functions. It's common to denote Schur functions in these variables as $S_{\lambda}\{p_k\}$, they depend only on a partition $\lambda$.

\subsubsection{Characters of the $sp_{2n}$}
The Lie algebra $sp_{2n}$ corresponds to the root system $C_n$. Characters of $sp_{2n}$ are symplectic Schur functions $ Sp_{\lambda}$ \cite{Schurs,JT}
\begin{equation}
   \chi_{\lambda}^{C_n} (x_i,\dots,x_n) = \frac{\text{det}\left(x_i^{\lambda_j+n-j+1}-x_i^{-(\lambda_j+n-j+1)}\right)}{\text{det}\left(x_i^{n-j+1}-x_i^{-(n-j+1)}\right)} = Sp_{\lambda}(x_i,\dots,x_n) .
\end{equation}
They also can be expressed via the $p_k$ variables, however in this case these variables are different:
\begin{equation}
    p_k = \sum_{i=1}^{n} \left(x_i^k +x_i^{-k} \right).
\end{equation}

\subsubsection{Characters of the $so_{n}$}
In the case of Lie algebra $so_n$, one has to distinguish between the cases of even and odd $n$. The root system $B_n$ corresponds to $so_{2n+1}$, and its characters can be obtained with the following formula

\begin{equation}
    \chi_{\lambda}^{B_n} (x_1, \dots,x_n) = \frac{\text{det}\left(x_i^{\lambda_j+n-j+1/2}-x_i^{-(\lambda_j+n-j+1/2)}\right)}{\text{det}\left(x_i^{n-j+1/2}-x_i^{-(n-j+1/2)}\right)}.
    \label{chB}
\end{equation}
In the case of $so_{2n}$ and the root system $D_n$, one gets
\begin{equation}
    \chi_{\lambda}^{D_n} (x_1, \dots,x_n) = \frac{\text{det}\left(x_i^{\lambda_j+n-j}+x_i^{-(\lambda_j+n-j)}-\delta_{j,n}\delta_{\lambda_n,0}\right)}{\text{det}\left(x_i^{n-j}+x_i^{-(n-j)}-\delta_{j,n}\right)},\ \ \text{when}\ \  \lambda_n = 0
    \label{chD0}
\end{equation}
and
\begin{equation}
    \chi_{\lambda}^{D_n} (x_1, \dots,x_n) =2\left( \frac{\text{det}\left(x_i^{\lambda_j+n-j}+x_i^{-(\lambda_j+n-j)}-\delta_{j,n}\delta_{\lambda_n,0}\right)}{\text{det}\left(x_i^{n-j}+x_i^{-(n-j)}-\delta_{j,n}\right)}-\frac{\text{det}\left(x_i^{\lambda_j+n-j}+x_i^{-(\lambda_j+n-j)}\right)}{\text{det}\left(x_i^{n-j}+x_i^{-(n-j)}\right)} \right),
    \label{chD1}
\end{equation}
when $\lambda_n \neq 0$.

These characters can be rewritten with orthogonal Schur functions $So_{\lambda}$. It's also convenient to express them via $p_k$ variables, however the two cases differ in the transition between $x_i$ and $p_k$ variables:
\begin{equation}
    \chi_{\lambda} ^{B_n} = So_{\lambda}\{p_k\},\quad p_k = \sum (x_i^k+x_i^{-k}) + 1,
\end{equation}
 \begin{equation}
    \chi_{\lambda} ^{D_n} = So_{\lambda}\{p_k\},\quad p_k = \sum (x_i^k+x_i^{-k}).
\end{equation}

\subsubsection{Schur functions}
To conclude we provide some expressions for the symplectic and orthogonal Schur functions via the ordinary Schur functions. They are enumerated with partitions $\Lambda = [\Lambda_1, \dots, \Lambda_n]$, however we omit the brackets $[\dots]$ to shorten the notation.
\begin{equation}
\begin{array}{l|l|l}
\text{Schur functions} & \text{Orthogonal Schurs functions} & \text{Symplectic Schur functions} \\
 S_1 & S_1 & S_1 \\
 S_2 & S_2-1 & S_2 \\
 S_{11} & S_{11} & S_{11}-1 \\
 S_3 & S_3-S_1 & S_3 \\
 S_{21} & S_{21}-S_1 & S_{21}-S_1 \\
 S_{111} & S_{111} & S_{111}-S_1 \\
 S_4 & S_4-S_2 & S_4 \\
 S_{31} & -S_2-S_{11}+S_{31}+1 & S_{31}-S_2 \\
 S_{22} & S_{22}-S_2 & S_{22}-S_{11} \\
 S_{211} & S_{211}-S_{11} & -S_2-S_{11}+S_{211}+1 \\
 S_{1111} & S_{1111} & S_{1111}-S_{11} \\
 S_5 & S_5-S_3 & S_5 \\
 S_{41} & S_1-S_3-S_{21}+S_{41} & S_{41}-S_3 \\
 S_{32} & S_1-S_3-S_{21}+S_{32} & S_{32}-S_{21} \\
 S_{311} & S_1-S_{21}-S_{111}+S_{311} & S_1-S_3-S_{21}+S_{311} \\
 S_{221} & S_{221}-S_{21} & S_1-S_{21}-S_{111}+S_{221} \\
 S_{2111} & S_{2111}-S_{111} & S_1-S_{21}-S_{111}+S_{2111} \\
 S_{11111} & S_{11111} & S_{11111}-S_{111} \\
\end{array}
\end{equation}
We stress once again that these relations are written for variables $p_k$, which differ for different root systems:
\begin{align}
   & A_n\, (sl(n+1)):  &&\!\!\!\!\!\!\!\!\!\!\!\!\!\!\!\!\!\!\!\!\!\!\!\! S\{p_k\},  && \!\!\!\!\!\!\!\!\!\!\!\!\!\!\!\!\!\!\!\!\!\!\!\! p_k = \sum x_i^k, \label{schurA}\\
   & B_n\, (so(2n+1)):  && \!\!\!\!\!\!\!\!\!\!\!\!\!\!\!\!\!\!\!\!\!\!\!\! So\{p_k\}, && \!\!\!\!\!\!\!\!\!\!\!\!\!\!\!\!\!\!\!\!\!\!\!\! p_k = \sum (x_i^k+x_i^{-k}) + 1, \label{schurB} \\
   & C_n\, (sp(2n)): && \!\!\!\!\!\!\!\!\!\!\!\!\!\!\!\!\!\!\!\!\!\!\!\! Sp\{p_k\}, && \!\!\!\!\!\!\!\!\!\!\!\!\!\!\!\!\!\!\!\!\!\!\!\!p_k = \sum (x_i^k+x_i^{-k}) , \label{schurC}\\
   & D_n\, (so(2n)): && \!\!\!\!\!\!\!\!\!\!\!\!\!\!\!\!\!\!\!\!\!\!\!\! So\{p_k\}, && \!\!\!\!\!\!\!\!\!\!\!\!\!\!\!\!\!\!\!\!\!\!\!\! p_k = \sum (x_i^k+x_i^{-k}). \label{schurD}
\end{align}

\section{Quantum and Macdonald dimensions \label{DimsSec}}
Quantum dimensions naturally arise in Chern-Simons theory as Wilson averages for unknotted loop and in knot theory as quantum invariants of the unknot.  Since the refinement of Chern-Simons theory is completed with the transition from Schur symmetric functions to Macdonald polynomials, we define Macdonald dimensions with Macdonald polynomials at a special point, analogously to quantum dimensions. In the Table \ref{alldimtable} we list the main notions of this section.

\subsection{Weyl vectors}
Quantum and Macdonald dimensions are defined with the help of Weyl vector and its variations:
\begin{align}
     \rho & = \frac{1}{2} \sum_{\alpha > 0} \alpha, \label{rho}\\
     \rho_k & = \frac{1}{2} \sum_{\alpha > 0} k_{\alpha} \, \alpha, \label{rhok} \\
     \rho_k^{*} & = \frac{1}{2} \sum_{\alpha > 0} k_{\alpha} \, \alpha_{*} = \frac{1}{2} \sum_{\alpha > 0} k_{\alpha} \, \alpha/u_{\alpha}, \label{rhostar}
\end{align}

\begin{align}
    r & = \frac{1}{2} \sum_{\alpha > 0} \alpha^{\vee}, \label{r}\\
    r_k & = \frac{1}{2} \sum_{\alpha > 0} k_{\alpha} \, \alpha^{\vee}, \label{rk} \\
    r_k^{*} & = \frac{1}{2} \sum_{\alpha > 0} k_{\alpha} \, \alpha^{*} = \frac{1}{2} \sum_{\alpha > 0} k_{\alpha} u_{\alpha}\, \alpha^{\vee}, \label{rstar}
\end{align}
where parameter $k_{\alpha}$ depends only on the length of the root $(\alpha,\alpha)$ and parameter $u_{\alpha}$ depends on the admissible pair $(R,S)$ and was defined in the previous section (\ref{uparm}).

\subsubsection{$A_n$}
Coordinates of Weyl vectors of root system $A_n$ in the orthogonal basis are the following:
\begin{equation}
    \begin{aligned}
        \left (\rho^{A_n} \right)_i=\left (r^{A_n} \right)_i &=n/2-(i-1), \\
        \left (\rho^{A_n}_k \right)_i= \left (r^{A_n}_k \right)_i & = kn/2-k(i-1). \\
    \end{aligned}
\end{equation}

\subsubsection{$B_n$}
Coordinates of Weyl vectors of root system $B_n$ in the orthogonal basis are the following:
\begin{equation}
    \begin{aligned}
        \left (\rho^{B_n} \right)_i &= (n-i)+1/2, \\
        \left (r^{B_n} \right)_i &= (n-i)+1,  \\
        \left (\rho^{B_n}_k \right)_i & = k(n-i)+k_s/2, \\
        \left (r^{B_n}_k \right)_i &= k(n-i)+k_s.  \\
    \end{aligned}
\end{equation}

\subsubsection{$C_n$}
Coordinates of Weyl vectors of root system $C_n$ in the orthogonal basis are the following:
\begin{equation}
    \begin{aligned}
        \left (\rho^{C_n} \right)_i &= (n-i)+1, \\
        \left (r^{C_n} \right)_i &= (n-i)+1/2,  \\
        \left (\rho^{C_n}_k \right)_i & = k(n-i)+k_l, \\
        \left (r^{C_n}_k \right)_i &= k(n-i)+k_l/2.  \\
    \end{aligned}
\end{equation}

\subsubsection{$D_n$}
Coordinates of Weyl vectors of root system $D_n$ in the orthogonal basis are the following:
\begin{equation}
    \begin{aligned}
        \left (\rho^{D_n} \right)_i=\left (r^{D_n} \right)_i &= (n-i), \\
        \left (\rho^{D_n}_k \right)_i= \left (r^{D_n}_k \right)_i & = k(n-i). \\
    \end{aligned}
\end{equation}

\subsection{Quantum and dual quantum dimensions}
In representation theory, quantum dimensions are the characters of irreducible representations of Lie algebras at the special point $x = q^{2\rho}$, where $\rho$ is the Weyl vector (\ref{rho}).
Characters of the Lie algebras $\chi^{R}_{\lambda}$ do factorize \cite[expr. 13.170]{DiFr} at the Weyl vector:
\begin{equation}\label{qd}
    \qD_{\lambda}^{R}:= \chi^{R}_{\lambda} \left(x = q^{2\rho} \right) =\prod_{\alpha\in R_+} \frac{[(\alpha,\lambda+\rho)]_q}{[(\alpha,\rho)]_q},
\end{equation}
where $(\cdot,\cdot)$ is the scalar product in the Euclidean space where the root system is embedded.

\bigskip

Characters of simple Lie algebras also factorize at another point, which we call the dual Weyl vector $r$ (\ref{r}), which is a half sum over all positive coroots. Hence, we define {dual quantum dimension} $\qDv_{\lambda}^{R}$ as a character at the point $q^{2r}$:
\be
     \qDv_{\lambda}^{R} : = \chi_{\lambda}^R\left(x  = q^{2r} \right)=
\prod_{\alpha \in R_+}\prod_{j=1}^{(\alpha^{\vee},\lambda)} \frac{[(\rho,\alpha^{\vee})+j]_q}{[(\rho,\alpha^{\vee})+j-1]_q}.
\label{qdv}
\ee

\subsection{Macdonald and dual Macdonald dimensions}
Analogously to quantum dimensions, we define their refined version based on Macdonald polynomials and call them Macdonald dimensions. Macdonald dimensions are Macdonald polynomials at the refined Weyl vector (\ref{rhok}):

\begin{equation}
    \MD_{\lambda}^{(R,S)} = P^{(R,S)}_{\lambda}(x = q^{2\rho_k}\,|\, t_{\alpha}^2\,|\, q^2, t^2).
\end{equation}

 Turns out that the Macdonald polynomials do not factorize at refined Weyl vector $\rho_k$,  but they do factorize and the factorization point is the dual refined Weyl vector $r^*_k$ (\ref{rstar}), that depends on parameters $k_\alpha$ and $u_{\alpha}$ \cite{Mac} (conj. 12.10):
\begin{equation}
\boxed{
   P^{(R,S)}_{\lambda}(x = q^{2r^*_k}\,|\, t_{\alpha}^2\,|\, q^2, t^2) =  \prod_{\alpha \in R_+} \, \prod_{j=1}^{(\alpha^{\vee},\lambda)}\,  \frac{\left\{t_{\alpha/2}\, t_{\alpha}\, q_{\alpha}^{(\rho_k,\alpha^{\vee})+j-1}\right\}}{\left\{ t_{\alpha/2}\, q_{\alpha}^{(\rho_k,\alpha^{\vee})+j-1}\right\}}, }
   \label{factor}
\end{equation}
where $t_{\alpha/2} = 1$ if $\alpha/2 \notin R$.

We call Macdonald polynomials in their factorization point dual Macdonald dimensions:
\begin{equation}
     \MDv_{\lambda}^{(R,S)} = P^{(R,S)}_{\lambda}(x = q^{2r_k^*}\,|\, t_{\alpha}^2\,|\, q^2, t^2).
\end{equation}
The distinction between $r_k$ and $r_k^*$ is only important when the admissible pair consists of two different root systems $R\neq S $. We discuss such dual Macdonald dimensions in section \ref{MixedMacSec}.

When $R = S$ we get $r_k^* = r_k$ and denote such dual Macdonald dimensions with one root system:

\begin{equation}
    \MDv_{\lambda}^{R} = P^{R}_{\lambda}(x = q^{2r_k}\,|\, t_{\alpha}^2\,|\, q^2, t^2) = \prod_{\alpha \in R_+} \, \prod_{j=1}^{(\alpha^{\vee},\lambda)}\,  \frac{\left\{t_{\alpha}\, q^{(\rho_k,\alpha^{\vee})+j-1}\right\}}{\left\{ q^{(\rho_k,\alpha^{\vee})+j-1}\right\}}.
    \label{MacDimDual}
\end{equation}

Refined Weyl vectors $\rho_k$ and $r_k$ do coincide for simply laced root systems, that is why Macdonald dimensions $\MD_{\lambda}^{A_n}$, $\MD_{\lambda}^{D_n}$ and $\MD_{\lambda}^{E_6}$, $\MD_{\lambda}^{E_7}$, $\MD_{\lambda}^{E_8}$ coincide with their dual versions, do factorize and can be universalized.

\subsection{On the root normalization}
We have already mentioned that Vogel parameters (Table \ref{vogelparm}) are written for the minimal normalization of roots, which means that the length of the longest root of the root system $\alpha_l$ equals to 2:
\begin{equation}
    (\alpha_l,\alpha_l) = 2.
\end{equation}
Roots that we used in calculations satisfy this requirement except for the root systems with one long root: $C_n$ and $G_2$.

In order to go to minimal normalization one can rescale the scalar product:
\begin{equation}
     (\cdot,\cdot) \rightarrow a(\cdot,\cdot),
     \label{scalarnorm}
\end{equation}
then
\begin{equation}
    \alpha \rightarrow \alpha, \quad \quad \rho \rightarrow \rho,  \quad\quad \lambda \rightarrow \lambda,
\end{equation}
\begin{equation}
    \alpha^{\vee} \rightarrow \frac{1}{a} \alpha^{\vee}, \quad\quad r \rightarrow \frac{1}{a} r
\end{equation}

We want to point out that dual versions of quantum and Macdonald dimensions $\qDv_{\lambda}^{R}$ and $\MDv_{\lambda}^{R}$ do not depend on the choice of normalization of roots:
\begin{equation}
    \qDv_{\lambda}^{R} \rightarrow \qDv_{\lambda}^{R}, \quad\quad \MDv_{\lambda}^{R} \rightarrow \MDv_{\lambda}^{R},
\end{equation}
but ordinary quantum and Macdonald dimensions do depend on the choice of normalization. For quantum dimensions one gets
\begin{equation}
    \qD_{\lambda}^{R} \rightarrow \prod_{\alpha\in R_+} \frac{[a(\alpha,\lambda+\rho)]_q}{[a(\alpha,\rho)]_q},
    \label{qdimrescale}
\end{equation}
which can be corrected with the rescaling of $q$: $ q \rightarrow q^{1/a}$. Expressions $X$ that are not in minimal normalization we denote with overline $\overline{X}$.

\subsection{Quantum and Macdonald dimensions in adjoint representation}
In this section, we provide explicit expressions for quantum dimensions $\qD_{\Adj}^{R}$, dual quantum dimensions $\qDv_{\Adj}^{R}$, dual Macdonald dimensions $\MDv_{\Adj}^{R}$ and some Macdonald dimensions $\MD_{\Adj}^{R}$ corresponding to adjoint representations of simple Lie algebras.

\subsubsection{$A_n$ $(sl_{n+1})$}
The highest weight of adjoint representation (in the orthogonal basis)
\begin{equation}
    \lambda_{\Adj}^{A_n} = [1,0,0,\dots,0,-1] =  \omega_1+\omega_n,
\end{equation}
 $\omega_1$ and  $\omega_n$ are the first and the last fundamental weights respectively.
It corresponds to the partition $[2,1^{n-1}]$ ($[2]$ for $sl(2)$, $[2,1]$ for $sl(3)$, $[2,1,1]$ for $sl(4)$ and so on).
\noindent
The quantum dimension in the adjoint representation coincides with the dual quantum dimension:
\begin{equation}
    \qD_{\Adj}^{A_n} = [n+2]_q [n]_q = \qDv_{\Adj}^{A_n}.
\end{equation}
The Macdonald dimension and the dual Macdonald dimension also coincide and in the adjoint representation are the following:
\begin{equation}
\boxed{
    \MD_{\Adj}^{A_n} =  \frac{\{t^{n+1}\}\{q\,t^{n+1}\}}{\{t^{n}\}\{q\,t^{n}\}} \prod_{j=2}^{n} \frac{\{t^j\}}{\{t^{j-1}\}} \frac{\{t^{n-j+2}\}}{\{t^{n-j+1}\}}= \frac{\{t^n\}\{t^{n+1}\}\{q\,t^{n+1}\}}{\{t\}^2\{q\,t^{n}\}}.}
    \label{MacDimA}
\end{equation}

\subsubsection{$B_n$ $(so_{2n+1})$}
The highest weight of adjoint representation (in the orthogonal basis)
\begin{equation}
    \lambda_{\Adj}^{B_n} = [1,1,0,0,\dots] = a \, \omega_2,
\end{equation}
where $a = 2$ for root system $B_2$ and $a = 1$ for all other $B_n$, $\omega_2$ is the second fundamental weight. It corresponds to the partition $[1,1]$.
\noindent
The quantum dimension and the dual quantum dimension in the adjoint representation:
\begin{equation}
    \qD_{\Adj}^{B_n} = \frac{[n+1/2]_q [2n]_q [2n-3]_q}{[n-3/2]_q[2]_q}, \quad  \qDv_{\Adj}^{B_n}  = \frac{[2n]_q[2n+1]_q}{[2]_q}.
\end{equation}
The dual Macdonald dimension in the adjoint representation:
\begin{equation}
\boxed{
    ^{\vee}\MD_{\Adj}^{B_n}=
    \frac{[n]_t[n-1]_t}{[2]_t}
    \frac{\{t^{2(n-2)}\,t_s^2\}\{q\,t^{2(n-2)}\,t_s^2\} \{t^{2(n-1)}\,t_s^2\} \{q\,t^{2(n-1)}\,t_s^2\} }{\{t^{n-2}\,t_s\}\{q\,t^{2(n-2)}\,t_s\}\{t^{n-1}\,t_s\}\{q\,t^{2n-3}\,t_s\}}.}
\end{equation}
The Macdonald dimension in this case looks more complicated and does not factorize:
\begin{equation}
    \begin{aligned}
&\MD_{\Adj}^{B_n}
= {\{t^n\}\over\{t\}}\left({\{t_s\}\over \{t\}}{\{t^{n-1}\}\over\{\xi_n\}} \{t_s\,t^{n-1}\}_{_+}\{q t^{n-2}\}_{_+}- {\{q\,t^{n-2}\}_{_+}\over\{t\}_{_+}} \frac{\{t^{n-1}\}\{t_s^2\,q^{-1}t^{-1}\}}{\{qt^{n-1}\}\{t_s\,t\xi_n\}}{_+}\right. \nn\\
&\left.{_+} \frac{\{t^{n-1}\}\{t_s\}\{t\,t_s\}\{t_s \xi_n\}}{\{t^2\}\{t\xi_n\}\{\xi_n\}\{t_st\xi_n\}}
\{q^2\,t^{2(n-1)}\}_{_+}\{q^2\,t^{2(n-2)}\}_{_+} - \frac{\{q\}}{\{q\,t^{n-1}\}}
-\frac{1}{2}\{t_s t^{n-1}\}_{_+}^2 -\frac{1}{2} \frac{\{t^n\}_{_+}}{\{t\}_{_+}} \{t_s^2 t^{2(n-1)}\}_{_+}\right),
     \label{cadjnonf}
    \end{aligned}
\end{equation}
where $\xi_n:=qt_st^{2(n-2)}$.

\subsubsection{$C_n $ $ (sp_{2n})$}
The highest weight of the adjoint representation (in the orthogonal basis)
\begin{equation}
    \lambda_{\Adj}^{C_n} = [2,0,\dots] = 2 \, \omega_1,
\end{equation}
where $\omega_1$ is the first fundamental weight.
It corresponds to the partition $[2]$.

The quantum dimension and the dual quantum dimension in the adjoint representation:
\begin{equation}
   \overline{\qD}^{C_n}_{\Adj} = \frac{[n]_q[2n+1]_q[2n+4]_q}{[n+2]_q[2]_q}, \quad \qDv_{\Adj}^{C_n} = \frac{[2n]_q[2n+1]_q}{[2]_q}.
\end{equation}
If one uses minimal normalization, the quantum dimensions of $C_n$ is the following (this formula is included in universal formula for quantum dimensions (\ref{uniQdim})):
\begin{equation}
  \qD^{C_n}_{\Adj} = \frac{[n/2]_q[n+1/2]_q[n+2]_q}{[(n+2)/2]_q[1/2]_q}.
\end{equation}
The dual Macdonald dimension in adjoint representation:
\begin{equation}
\boxed{
    ^{\vee}\MD_{\Adj}^{C_n} = \frac{\{t^n\}\{q\,t^n\} \{t^{2(n-1)}\,t_l^2\}\{q\,t^{2(n-1)}\,t_l^2\}}{\{t\}\{q\,t\} \{t^{n-1}\,t_l\}\{q\,t^{n-1}\,t_l\}}.}
\end{equation}
And Macdonald dimension (overline means that this answer is not in the minimal normalization):
\begin{equation}
    \overline{\MD}^{C_n}_{Adj}=  P^{C_n}_{Adj}\left(x = q^{2\rho_k}\,|\,t_l^2\,|\,q^2,t^2\right) =
    [n]_t{\{q t^n\}\over\{q t\}}\left({\{t_l^3\xi_n^3\} \over \{t_l\xi_n\}}+{\{q^{-1} \xi_n\}\over \{q\xi_n\}}\right), \quad \xi_n = t^{n-1} t_l.
\end{equation}

We list more examples of Macdonald dimensions in Appendix C.

\subsubsection{$D_n$ $(so_{2n})$}
The highest weight of adjoint representation (in the orthogonal basis)
\begin{equation}
    \lambda_{\Adj}^{D_n} = [1,1,0,0,\dots],
\end{equation}
it corresponds to Young diagram $[1,1]$.
Quantum dimension in adjoint representation coincides with dual quantum dimension:
\begin{equation}
    \qD_{\Adj}^{D_n} = \frac{[n]_q [2n-1]_q [2n-4]_q}{[n-2]_q[2]_q} =  \qDv_{\Adj}^{D_n}.
\end{equation}
and the Macdonald dimension in the adjoint representation, which coincides with the dual Macdonald dimension:
\begin{equation}
\boxed{
    \MD_{Adj}^{D_n} = \frac{ \{t^n\} \{t^{2 n-4}\} \{t^{2
   n-2}\} \{q\, t^{2 n-2}\}}{\{t\} \{t^2\}
  \{t^{n-2}\} \{q\, t^{2 n-3}\}}.}
  \label{MacDimD}
\end{equation}

\subsubsection{$E_6$}
\begin{equation}
    \qD_{\Adj}^{E_6} = \frac{[8]_q[9]_q[13]_q}{[3]_q[4]_q} = \qDv_{\Adj}^{E_6},
\end{equation}

\begin{equation}
    \MD^{E_6}_{\Adj} = \frac{\{t^8\}
   \{t^9\}
  \{t^{12}\} \{q\,
   t^{12}\}}{\{t\}
   \{t^3\} \{t^4\}
  \{q \,t^{11}\}}.
  \label{MacDimE6}
\end{equation}

\subsubsection{$E_7$}

\begin{equation}
     \qD_{\Adj}^{E_7} = \frac{[12]_q[14]_q[19]_q}{[4]_q[6]_q}= \qDv_{\Adj}^{E_7},
\end{equation}

\begin{equation}
    \MD_{\Adj}^{E_7} =  \frac{\{t^{12}\} \{t^{14}\} \{t^{18}\}
   \{q\, t^{18}\}}{\{t\} \{t^4\}
   \{t^6\} \{q\, t^{17}\}}.
   \label{MacDimE7}
\end{equation}

\subsubsection{$E_8$}

\begin{equation}
     \qD_{\Adj}^{E_8} = \frac{[20]_q[24]_q[31]_q}{[6]_q[10]_q}=\qDv_{\Adj}^{E_8},
\end{equation}

\begin{equation}
     \MD_{\Adj}^{E_8} = \frac{\{t^{20}\}\{t^{24}\}\{t^{30}\}
   \{q\, t^{30}\}}{\{t\}\{t^6\}
  \{t^{10}\} \{q\, t^{29}\}}.
  \label{MacDimE8}
\end{equation}

\subsubsection{$F_4$}

\begin{equation}
\qD_{\Adj}^{F_4} = \frac{[10]_q[13/2]_q[6]_q}{[3]_q[5/2]_q}, \quad \qDv_{\Adj}^{F_4}  = \frac{[8]_q[12]_q[13]_q}{[4]_q[6]_q},
\end{equation}

\begin{equation}
    ^{\vee}\MD_{\Adj}^{F_4} = \frac{\{t^3\} \{t^4 t_s^2\} \{t^4
   t_s^4\}\{t^6 t_s^6\} \{q\, t^4
   t_s^4\} \{q \, t^6 t_s^6\}}{\{t\}  \{t^2 t_s\} \{t^2
   t_s^2\} \{t^3 t_s^3\} \{q\, t^4
   t_s\} \{q \,t^5 t_s^3\}}.
\end{equation}

\subsubsection{$G_2$}
\begin{equation}
    \overline{\qD}_{\Adj}^{G_2} = \frac{[8]_q[7]_q[15]_q}{[3]_q[4]_q[5]_q}, \quad \qDv_{\Adj}^{G_2} = \frac{[7]_q[8]_q}{[4]_q}.
\end{equation}
In minimal normalization the quantum dimension of $G_2$ is the following (this formula is included in universal formula for quantum dimensions (\ref{uniQdim})):
\begin{equation}
    \qD_{\Adj}^{G_2} = \frac{[8/3]_q[7/3]_q[5]_q}{[4/3]_q[5/3]_q}.
\end{equation}
The dual Macdonald dimension:
\begin{equation}
   ^{\vee}\MD^{G_2}_{\Adj} = \frac{\{t_3^2\} \{t^3 t_3^3\} \{q\,
   t^3 t_3^3\} \{q^2\, t^3 t_3^3\}}{
   \{t_3\} \{t\, t_3\} \{q \, t\,
   t_3^2\} \{q^2\, t \,t_3^3\}}.
\end{equation}

\section{Mixed Macdonald polynomials and their dimensions \label{MixedMacSec}}
In this section we consider Macdonald polynomials associated with two different root systems $(R,S)$. We discuss parameters of these polynomials and their factorization.

\subsection{Factorization formula}
Macdonald polynomials depend on a pair of root systems $(R,S)$. When $R\neq S$ we get different $q_{\alpha}$:
\begin{equation}
    q_{\alpha} = q^{u_{\alpha}},
\end{equation}
where for each $\alpha \in R$ there exists a unique $u_\alpha \in R_+$ such that
\begin{equation}
    \alpha_{*}:=\alpha/u_{\alpha} \in S, \,\, \alpha^* = (\alpha_*)^{\vee} = u_{\alpha} \alpha^{\vee}.
\end{equation}
The factorization of Macdonald polynomials associated with root systems in the case of general admissible pair $(R,S)$ occurs at the variation of the Weyl vector that involves parameters $k_{\alpha}$ and $u_{\alpha}$:
\begin{equation}
 r_k^{*} = \frac{1}{2} \sum_{\alpha > 0} k_{\alpha} \alpha^{*} = \frac{1}{2} \sum_{\alpha > 0} k_{\alpha} u_{\alpha} \alpha^{\vee}.
\end{equation}
The formula for Macdonald factorization is the following:
\begin{equation}
   P^{(R,S)}_{\lambda}\left(x = q^{2r^*_k}\,|\, t_{\alpha}^2\,|\, q^2, t^2 \right) =  \prod_{\alpha \in R_+} \, \prod_{j=1}^{(\alpha^{\vee},\lambda)}\,  \frac{\left\{t_{\alpha/2}\, t_{\alpha}\, q_{\alpha}^{(\rho_k,\alpha^{\vee})+j-1}\right\}}{\left\{t_{\alpha/2}\, q_{\alpha}^{(\rho_k,\alpha^{\vee})+j-1}\right\}}.
\end{equation}

\subsection{$(B_n,C_n)$}

There are two clusters of positive roots (\ref{rootsBplus}) in the case of the root system $B_n$: $|\alpha|^2=| \varepsilon \pm \varepsilon|^2 = 2$ and $|\alpha|^2= |\varepsilon|^2 = 1$, hence
\begin{equation}
    \begin{array}{llll}
   u_{|\alpha|^2 = 2} = 1, & k_{|\alpha|^2 = 2} = k, &    q_{|\alpha|^2 = 2} = q, & t_{|\alpha|^2 = 2} =t= q^{k},  \\
      u_{|\alpha|^2 = 1} = 1/2, & k_{|\alpha|^2 = 1} = k_s, &    q_{|\alpha|^2 = 1} = q^{1/2},&  t_{|\alpha|^2 = 1} = t_s=q^{k_s/2}, \\
      &&& t_{\alpha/2} = 1 \,\, \forall \,\, \alpha \in B_n.
    \end{array}
\end{equation}

\begin{equation}
    r_k^* = k (n-i)+k_s/2
\end{equation}
and factorization occurs at the point
\begin{equation}
    q^{2r_k^*}: \quad x_i = t^{2(n-i)}\, t_s^2.
\end{equation}
In the representation $\Adj^{B_n} = [1,1]$ the dual Macdonald dimension is the following:
\begin{align}
& \MDv_{\Adj}^{(B_n,C_n)}=P_{\Adj}^{(B_n,C_n)}\left(x = q^{2r_k^*}\,|\, t_s^2\,|\,q^2,t^2 \right) =  \\
&  \nn \quad\quad\quad
    \frac{[n]_t[n-1]_t}{[2]_t}
    \frac{\{t^{2(n-2)}\,t_s^2 \}\{t^{2(n-1)}\,t_s^2 \}\{q\,t^{2(n-1)}\,t_s^2 \}\{q^{1/2}\,t^{n-1}\,t_s^2 \}\{q^{1/2}\,t^{n-2}\,t_s^2 \}}{\{t^{n-2}\, t_s\}\{ t^{n-1}\, t_s\}\{ q\,t^{2n-3}\, t_s\}\{ q^{1/2}\,t^{n-1}\, t_s\}\{ q^{1/2}\,t^{n-2}\, t_s\}}.
\end{align}
 When $k=k_s=1$ ($t=q$ and $t_s=q^{1/2}$) $^{\vee}\MD_{\Adj}^{(B_n,C_n)}$ goes to
 \begin{equation}
     ^{\vee}\MD_{\Adj}^{(B_n,C_n)}\,\xrightarrow[]{t\rightarrow q,\, t_s \rightarrow q^{1/2}} \, \frac{[2n]_q[2n-3]_q [n+1/2]_q}{[2]_q[n-3/2]_q},
\end{equation}
when $t=q$ and $t_s=q$ $ \MDv_{\Adj}^{(B_n,C_n)}$ goes to
\begin{equation}
    \MDv_{\Adj}^{(B_n,C_n)}\,\xrightarrow[]{t\rightarrow q,\, t_s \rightarrow q} \, \frac{[2(n-1)]_q[2n+1]_q [n+3/2]_q}{[2]_q[n-1/2]_q}.
\end{equation}

\subsection{$(C_n, B_n)$}
There are two groups of positive roots (\ref{rootsCplus}) in the case of the root system $C_n$: $|\alpha|^2=|\varepsilon \pm \varepsilon|^2 = 2$ and $|\alpha|^2= |2\varepsilon|^2 = 4$.
\begin{equation}
    \begin{array}{llll}
   u_{|\alpha|^2 = 2} = 1, &  k_{|\alpha|^2 = 2} = k,  &    q_{|\alpha|^2 = 2} = q, & t_{|\alpha|^2 = 2} =t= q^{k}  \\
      u_{|\alpha|^2 = 4} = 2, & k_{|\alpha|^2 = 4} = k_l &    q_{|\alpha|^2 = 4} = q^{2}&  t_{|\alpha|^2 = 4} = t_l=q^{2 k_l} \\
      &&& t_{\alpha/2}^2 = 1 \,\, \forall \,\, \alpha \in C_n.\\
    \end{array}
\end{equation}
\begin{equation}
    r_k^* = k(n-i) +k_l.
\end{equation}
The factorization occurs at the point (which exactly coincides with factorization point for polynomials $P^{C_n}_{\lambda}$):
\begin{equation}
    q^{2r^*_k}: \quad x_i = t^{2(n-i)}\, t_l.
\end{equation}
In the adjoint representation $\Adj^{C_n} = [2]$:
\begin{equation}
  \MDv_{\Adj}^{(C_n,B_n)} =  P_{\Adj}^{(C_n,B_n)} \left(x = q^{2r_k^*}\,|\, t_l^2\,|\,q^2,t^2 \right) = [n]_t \frac{\{q\, t^n\}}{\{q \,t\}}\frac{\{q \,t^{2(n-1)}\, t_l^2\}}{\{q\,t^{(n-1)}\,t_l\}} \frac{\{t^{2(n-1)}\, t_l^2\}}{\{\,t^{n-1}\,t_l\}},
\end{equation}
which goes to
\begin{equation}
    \MDv_{\Adj}^{(C_n,B_n)} \,\xrightarrow[]{t\rightarrow q,\, t_l \rightarrow q^2} \, \frac{[2n+3]_q[2n+2]_q[n]_q}{[2]_q[n+2]_q}
\end{equation}
and
\begin{equation}
    \MDv_{\Adj}^{(C_n,B_n)} \,\xrightarrow[]{t\rightarrow q,\, t_l \rightarrow q} \, \frac{[2n]_q[2n+1]_q}{[2]_q}.
\end{equation}
\noindent
In representation $[1,1]$ factorization is the following:
\begin{align}
\MDv_{[1,1]}^{(C_n,B_n)}=P_{[1,1]}^{(C_n,B_n)}\left(x = q^{2r_k^*}\,|\, t_l^2\,|\,q^2,t^2 \right) =
 \frac{[n]_t[n-1]_t}{[2]_t}
\frac{\{t^{2(n-2)}t_l^2\}}{\{t^{n-2}t_l\}}
\frac{\{q\,t^{2(n-1)}t_l\}}{\{t^{n-1}t_l\}} \frac{\{t^{2(n-1)}t_l^2\}}{\{q\,t^{2n-3}t_l\}}.
\end{align}
 When $t=q$ and $t_l=q^2$ $ \MDv_{[1,1]}^{(C_n,B_n)}$ goes to
\begin{equation}
    \MDv_{[1,1]}^{(C_n,B_n)}\,\xrightarrow[]{t\rightarrow q,\, t_l \rightarrow q^2} \, \frac{[2n+2]_q[2n+1]_q[n-1]_q}{[2]_q[n+1]_q}
\end{equation}
and for $t=q$ and $t_l=q$
 \begin{equation}
    \MDv_{[1,1]}^{(C_n,B_n)} \,\xrightarrow[]{t\rightarrow q,\, t_l \rightarrow q} \, \frac{[2n]^2_q[2(n-1)]_q}{[2]_q[2n-1]_q}.
\end{equation}

\subsection{$(BC_n, B_n)$}
The root system $BC_n$ is not reduced and has the following positive roots:
\begin{equation}
        2\varepsilon_i, \quad \varepsilon_i, \quad \varepsilon_i+\varepsilon_j, \quad \varepsilon_i-\varepsilon_j, \quad \text{where}\quad i,j = \overline{1,n}, \quad i<j,
\end{equation}
so it is a combination of roots of the root systems $B_n$ and $C_n$.

There are three groups of roots in the case of the root system $BC_n$: $|\alpha|^2=|\varepsilon \pm \varepsilon|^2 = 2$,  $|\alpha|^2= |2\varepsilon|^2 = 4$ and  $|\alpha|^2= |\varepsilon|^2 = 1$.  In the case of the admissible pair $(BC_n, B_n)$ the Macdonald parameters are the following:
\begin{equation}
    \begin{array}{lllll}
   u_{|\alpha|^2 = 2} = 1, &  k_{|\alpha|^2 = 2} = k,  &    q_{|\alpha|^2 = 2} = q, & t_{|\alpha|^2 = 2} =t= q^k , & t_{(\varepsilon\pm\varepsilon)/2} = 1 \\
      u_{|\alpha|^2 = 4} = 2, & k_{|\alpha|^2 = 4} = k_l &    q_{|\alpha|^2 = 4} = q^{2}&  t_{|\alpha|^2 = 4} = b =q^{2 k_l} & t_{2\varepsilon/2} = a,\\
       u_{|\alpha|^2 = 1} = 1, & k_{|\alpha|^2 = 1} = k_s, &    q_{|\alpha|^2 = 1} = q,&  t_{|\alpha|^2 = 1} =  a =q^{k_s}, & t_{\ve/2} = 1. \\
    \end{array}
\end{equation}
We use $a$ and $b$ instead of $t_s$ and $t_l$ as it was done in Koornwinder's original work \cite{Koorn}.
\begin{equation}
    r_k^* = k(n-i) +k_l+k_s
\end{equation}
The factorization occurs at the point:
\begin{equation}
    q^{2r^*_k}: \quad x_i = t^{2(n-i)}\, a^2 \, b.
\end{equation}
In representation $[1,1]$:
\begin{align}
    \MDv_{[1,1]}^{(BC_n,B_n)}= &   P_{[1,1]}^{(BC_n,B_n)} \left(x = q^{2r_k^*}\,|\, a^2, b^2\,|\,q^2,t^2 \right)= \\ & \nn [n]_t[n-1]_t [2]_t^{-1} \frac{\{a^2 \,b^2\, t^{2(n-2)}\} \{a^2\, b^2\, t^{2(n-1)}\}}{\{a \,b\, t^{n-2}\} \{a\, b\, t^{n-1}\}} \frac{
   \{q \, a^2\, b\, t^{2 (n-2)}\} \{q \,a^2\, b \, t^{2
   (n-1)}\}}{\{q \,a \,b \, t^{2 (n-2)}\} \{q \, a \, b \,
   t^{2 n-3}\}}.
\end{align}
It can be reduced to the following expressions:
 \begin{equation}
    \MDv_{[1,1]}^{(BC_n,B_n)} \,\xrightarrow[]{t\rightarrow q,\, a \rightarrow q,\, b \rightarrow q^2} \, \frac{[2n+4]_q[2n+3]_q[2n+2]_q[n]_q[n-1]_q}{[2n]_q[n+2]_q[n+1]_q[2]_q},
\end{equation}
 \begin{equation}
    \MDv_{[1,1]}^{(BC_n,B_n)} \,\xrightarrow[]{t\rightarrow q,\, a \rightarrow q,\, b \rightarrow q} \, \frac{[2n+2]^2_q[2n]_q[n-1]_q}{[2n-1]_q[n+1]_q[2]_q}.
\end{equation}

\subsection{$(BC_n, C_n)$}
In the case of the admissible pair $(BC_n, C_n)$ the Macdonald parameters are the following:
\begin{equation}
    \begin{array}{lllll}
   u_{|\alpha|^2 = 2} = 1, &  k_{|\alpha|^2 = 2} = k,  &    q_{|\alpha|^2 = 2} = q, & t_{|\alpha|^2 = 2} =t= q^k , & t_{(\varepsilon\pm\varepsilon)/2} = 1 \\
      u_{|\alpha|^2 = 4} = 1, & k_{|\alpha|^2 = 4} = k_l &    q_{|\alpha|^2 = 4} = q&  t_{|\alpha|^2 = 4} = b=q^{k_l} & t_{2\varepsilon/2} = a,\\
       u_{|\alpha|^2 = 1} = 1/2, & k_{|\alpha|^2 = 1} = k_s, &    q_{|\alpha|^2 = 1} = q^{1/2},&  t_{|\alpha|^2 = 1} = a=q^{k_s/2}, & t_{\ve/2} = 1. \\
    \end{array}
\end{equation}
\begin{equation}
    r_k^* = k(n-i) +k_l/2+k_s/2.
\end{equation}
The factorization occurs at the point (which is exactly the same as in the case of $(BC_n,B_n)$):
\begin{equation}
    q^{2r^*_k}: \quad x_i = t^{2(n-i)}\, a^2 \, b.
\end{equation}
In representation $[1,1]$:
\begin{align}
 \MDv_{[1,1]}^{(BC_n,C_n)} = & P_{[1,1]}^{(BC_n,C_n)} \left(x = q^{2r_k^*}\,|\, a^2, b^2\,|\,q^2,t^2 \right) = \\ \nn &   [n]_t[n-1]_t[2]_t^{-1} \frac{\{a^2 \,b^2\, t^{2(n-2)}\} \{a^2\, b^2\, t^{2(n-1)}\} \{q\,a^2\,b^2\,t^{2(n-1)}\}}{\{a \,b\, t^{n-2}\} \{a\, b\, t^{n-1}\}\{q^{1/2}\,a\,b\,t^{n-1}\}} \frac{
   \{q^{1/2}\,a^2\, b\, t^{n-2}\} \{q^{1/2}\,a^2 \,b\, t^{n-1}\}}{
   \{q^{1/2 } \,a \, b\, t^{n-2}\} \{q\,a^2\, b^2 \,t^{2
   n-3}\}}.
\end{align}
It can be reduced to the following expressions:
 \begin{equation}
    \MDv_{[1,1]}^{(BC_n,C_n)} \,\xrightarrow[]{t\rightarrow q,\, a \rightarrow q^{1/2},\, b \rightarrow q} \, \frac{[2n+2]_q[2n-1]_q[n+3/2]_q[n-1]_q}{[n+1]_q[n-1/2]_q[2]_q},
\end{equation}
 \begin{equation}
      \MDv_{[1,1]}^{(BC_n,C_n)} \,\xrightarrow[]{t\rightarrow q,\, a \rightarrow q,\, b \rightarrow q} \, \frac{[2n+3]_q[2n]_q[n+5/2]_q[n-1]_q}{[n+1]_q[n+1/2]_q[2]_q}.
\end{equation}

\section{Universality in Chern-Simons theory and its refinement \label{UniSec}}

In this section we start with the simplest universality formula --- the universal dimension formula for adjoint representations --- and see how it transforms into the universal quantum dimension formula and finally to universal refined quantum dimension formula --- universal Macdonald dimension. However, universality in the refined case holds only for Lie algebras associated with simply laced root systems.

Parallel to quantum and Macdonald dimensions exist their dual versions: dual quantum and dual Macdonald dimensions. We discuss the possibility of universalization in case of dual quantum dimensions. Naturally the dual Macdonald dimensions also universalize in the simply-laced case, because in this case they coincide with Macdonald dimensions.

\subsection{\label{UniCS}Universality in Chern-Simons theory. Quantum dimensions and dual quantum dimensions}

The first universal formula is the formula for dimensions of adjoint representations \cite{Vogel95} for all simple Lie algebras:
\begin{equation}
\boxed{
  D_{\text{Adj}}(\mathfrak{a},\mathfrak{b},\mathfrak{c})  = {(\mathfrak{a} - 2\mathfrak{t})(\mathfrak{b} - 2\mathfrak{t})(\mathfrak{c} - 2\mathfrak{t})\over\mathfrak{a}\mathfrak{b}\mathfrak{c}}.
  }
\end{equation}

\vspace{10pt}
In Chern-Simons theory the dimension of representations transforms into the {\bf quantum dimension} $\qD_{\lambda}$, which is the Wilson average of unknotted loop (colored HOMFLY-PT invariant of the unknot in knot theory). In representation theory, quantum dimension is a character of the Lie algebras at the Weyl vector. Quantum dimensions (\ref{qd}) were universalized in the adjoint representation \cite{Westbury03, MkrtQDims}:
\begin{align}
\boxed{
        \qD_{\text{Adj}}(\mathfrak{a},\mathfrak{b},\mathfrak{c}) ={[\mathfrak{a}/2 - \mathfrak{t}]_q[\mathfrak{b}/2 - \mathfrak{t}]_q[\mathfrak{c}/2 - \mathfrak{t}]_q\over[\mathfrak{a}/2]_q[\mathfrak{b}/2]_q[\mathfrak{c}/2]_q}.
        }
 \label{uniQdim}
    \end{align}

\bigskip

Characters of simple Lie algebras also factorize at another point, which we call the dual Weyl vector $r$. We defined {\bf dual quantum dimension} $\qDv_{\lambda}^{R}$ as a character at the point $q^{2r}$ (\ref{qdv}).
In the limit of $q\to 1$, both quantum dimensions and dual quantum dimensions certainly reduce to the ordinary dimensions. In the case of simply laced root systems ($A_n$, $D_n$, $E_6$, $E_7$, $E_8$), the Weyl vector and the dual Weyl vector coincide $\rho = r$, so the quantum and the dual quantum dimensions coincide as well.

The question is whether or not  the dual quantum dimensions can be universalized alongside with the quantum dimensions. It meets, at least, two serious problems. First of all, if there is a universal formula for the dual quantum dimensions, there should exist at once two universal formulas coinciding for the simply laced root systems and differing for the non-simply-laced ones. It is quite an intricate requirement, which looks impossible to satisfy. Indeed, the difference of these two putatively existing universality formulas, $\Delta=\qDv_{\text{Adj}}(\mathfrak{a},\mathfrak{b},\mathfrak{c})-\qD_{\text{Adj}}(\mathfrak{a},\mathfrak{b},\mathfrak{c})$ has to vanish at the whole line associated with the root system of type $D$. This line is given by $2\mathfrak{a}+\mathfrak{b}=0$, i.e. the difference $\Delta$ has to be proportional to $2\mathfrak{a}+\mathfrak{b}$. However, the root system of type $B$ is also associated with this line, where $\Delta$ vanishes. At the same time, $\qDv_{\text{Adj}}^{B_n}-\qD_{\text{Adj}}^{B_n}\ne 0$!\footnote{One also can find a discussion of uniqueness of the universal formula for the quantum dimensions in \cite{AM21}.} Secondly, the dual quantum dimensions for the root systems $B_n$ and $C_n$ coincide in adjoint representation (which is a corollary of duality between the symplectic and orthogonal groups \cite{dual1,dual2,dual3,dual4}) though they are associated with completely distinct Vogel's parameters with even distinct dependence on the rank $n$. This condition is also not that simple to satisfy.

\subsection{Universality in refined Chern-Simons theory. Macdonald dimensions and dual Macdonald dimensions \label{UniRefCS}}

Now we take the next step and study universal quantities \cite{KS} in the refined Chern-Simons theory \cite{AgSh1,AgSh2,AM1,R,AvMkrtString,Mane,AM2,AM3}. In this case, there is a new type of dimension: the refined version of the quantum dimension and the dual quantum dimension, which we call Macdonald dimension and dual Macdonald dimension accordingly. In order to refine Chern-Simons theory with $SU(N)$ gauge group, one should go from the $sl_N$-characters, the Schur functions, to Macdonald polynomials  \cite{AgSh1, AgSh2}. Hence, it seems natural to base these refined dimensions on the Macdonald polynomials $P^{(R,S)}_{\lambda}(x\,|\,t_{\alpha}^2\,|\,\,q^2,t^2)$ associated with different root systems $R$, which we discussed in great detail in the previous sections.

Similarly to quantum dimension and dual quantum dimension, one naturally defines { Macdonald dimension} $\MD^R_{\lambda}$ and dual Macdonald dimension $\dMD_{\lambda}^{R}$ with the refined Weyl vector $\rho_k$ (\ref{rhok}) and the dual refined Weyl vector $r_k$ (\ref{rk}):
\begin{align}
    \MD^R_{\lambda}:= &P^{R}_{\lambda}\left(x = q^{2\rho_k}\,|\,t_\alpha^2\,|\,q^2,t^2\right), \label{MD}\\
    \dMD_{\lambda}^{R} := &P^{R}_{\lambda}\left(x = q^{2r_k}\,|\, t_{\alpha}^2\,|\, q^2, t^2 \right) \label{MDv}.
\end{align}
with the refined Weyl vector $\rho_k$ (\ref{rhok}). Here we discuss only  polynomials $P^{R}_{\lambda} (x\,|\,t_{\alpha}^2\,|\,\,q^2,t^2)$ that depend on one root system $(R,R)$.

Macdonald polynomials $ P^{R}_{\lambda}(x\,|\, q^2, t^2)$ associated with the simply laced root systems $R = A_n, D_n, E_6, E_7, E_8$  depend only on two parameters $q^2$ and $t^2$ and factorize at the refined Weyl vector $\rho_k$ (\ref{rhok}), since it coincides with the dual refined Weyl vector $r_k$ (\ref{rk}). Hence the Macdonald dimensions (\ref{MD}) in this case coincide with the dual Macdonald dimensions (\ref{MDv}).

In the case of non-simply-laced root systems $R = B_n,C_n,F_4,G_2$, the Macdonald polynomials $P^{R}_{\lambda}(x\,|\,t_{\alpha}^2\,|\,q^2,t^2)$  depend on an additional parameter $t_{\alpha}^2$ associated with the root of a distinct length.  Since, in the non-simply-laced case, there are several deformation parameters $t^2$, $t_{\alpha}^2$, and, in the simply laced case, there is just one $t^2$, one does not have to expect that there is a universality in the generic case. However, in the simply-laced case, the universality still persists.

There is another argument against universality of all algebras after the refinement. It is connected to the fact that the quantum dimensions celebrate universality, while the dual quantum dimensions seem to not universalize as we discussed in the previous section. Hence, it is expected that their refined versions do not universalize as well.

At the same time, the adjoint Macdonald polynomials associated with the simply laced root systems factorize according to formulas  (\ref{MacDimA}), (\ref{MacDimD}), (\ref{MacDimE6}), (\ref{MacDimE7}) and  (\ref{MacDimE8}) and can be unified with a universal formula which one can call {\bf the simply laced universal Macdonald dimension} (which coincides with the dual one):

\begin{equation}\label{main1}
\boxed{
    \MD_{\text{Adj}}(\mathfrak{a},\mathfrak{b},\mathfrak{c}) =    - 
    {\{t^{\mathfrak{a}+\mathfrak{b}/2+\mathfrak{c}}\}\{t^{\mathfrak{a}+\mathfrak{b}+\mathfrak{c}/2}\}
    \{t^{\mathfrak{a}+\mathfrak{b}+\mathfrak{c}}\}\{q^{-\mathfrak{a}/2}\,t^{\mathfrak{a}+\mathfrak{b}+\mathfrak{c}}\} \{q\}\over
    \{q^{\mathfrak{a}/2}\}\{t^{\mathfrak{b}/2}\}
    \{t^{\mathfrak{c}/2}\}\{q\,t^{\mathfrak{a}+\mathfrak{b}+\mathfrak{c}-1}\}\{t\}}.
    }
\end{equation}
Let us discuss some properties of this formula. First of all, it is symmetric under permutations of $u=q^{\mathfrak{a}}$, $v=t^{\mathfrak{b}}$, $w=t^{\mathfrak{c}}$, and not of the Vogel's parameters $\mathfrak{a}$, $\mathfrak{b}$, $\mathfrak{c}$: it follows from the analyses of the refined universality in \cite{BMM} that these are $u$, $v$ and $w$ that are the proper universal parameters in the refined case.  This particular form for the universal Macdonald dimension coincides with (26) in \cite{BMM}. At $t=q$, it coincides with formula for the universal quantum dimension (\ref{uniQdim}). 

Second, we also point out that, similarly to the quantum dimensions, the dual Macdonald dimensions (\ref{MacDimDual}) and the universal formula (\ref{main1}) do not depend on the choice of the scalar product normalization (\ref{scalarnorm}). In the unrefined case, changing the normalization is associated with simultaneous rescaling of all Vogel's parameters, and, in the case of quantum dimensions, this rescaling is made with a simultaneous changing of $q$, which leaves  (\ref{uniQdim}) intact. In the refined case, changing the scalar product normalization is not given simply by a rescaling of parameters $\mathfrak{a}$, $\mathfrak{b}$, $\mathfrak{c}$, and formula (\ref{main1}), though being invariant under changing the normalization, is not invariant under simultaneous changing the Vogel parameters and the parameters $q$ and $t$.

One may say that the universal refined formulas depend on five quantities: $u$, $v$, $w$, $q$ and $t$, and $u$, $v$, $w$ are Vogel's parameters invariant under changing the normalization of the scalar product.

The universality formula (\ref{main1}) for the simply laced algebras is \textbf{the main result} of the paper.
It is in accordance with earlier claims by authors of \cite{KS,AM1,Mane}, who studied the refinement of the Chern-Simons partition function on $S^3$ and also realized that universality holds for the simply laced algebras.

\section{Conclusion}

The goal of this paper is two-fold. First of all, we introduce the Macdonald and dual Macdonald dimensions, which are counterparts of the quantum dimensions and dual quantum dimensions. This requires an accurate description of the Macdonald polynomials associated with arbitrary root systems within Macdonald-Cherednik theory. The second goal is to use the introduced dimensions in order to demonstrate that the Vogel's universality can be extended from the adjoint quantum dimensions to the adjoint Macdonald dimensions in the case of simply laced root systems.

The Macdonald dimensions are associated with the refinement of Chern-Simons theory \cite{AgSh1,AgSh2}, thus, our results expose the fact (which was earlier also observed in \cite{KS, AM1, Mane}) that the Vogel's universality in the refined Chern-Simons theory involves only the simply laced root systems. We explain the origin of this phenomenon: it turns out that only the dual Macdonald dimensions factorize, and the Macdonald dimensions do not (in variance with the quantum dimensions, which factorize), but, in the simply laced case, the both types of Macdonald dimensions coincide, and, hence, factorize, and admit universal description in the adjoint case.

This description was extended to the square of the adjoint polynomials in \cite{BMM}, and, by now, our understanding of the refined Vogel's universality is restricted to these cases. It would be very interesting to study the possibility of universal description of more quantities, and to make the status of the refined Vogel's universality more clear.

\section*{Acknowledgments}
The author is very grateful to Andrei Mironov for valuable discussions and help with the preparation of this paper. The author also thanks Alexei Morozov, Aleksandr Popolitov, Nikita Tselousov and the anonymous reviewer for valuable comments and interest in the topic. The author is also grateful to the organizers of the Workshop ``Universal description of Lie algebras, Vogel theory, applications" in Dubna (April, 2025).
The work was supported by the Foundation for the Advancement of Theoretical Physics and Mathematics ``BASIS" (grant no. 23-1-1-48-2).

\section*{Appendix A. Macdonald densities}

In this section we use the notation of I.\,Macdonald paper \cite{Mac} and the parameters are $q$, $t$, $t_{\alpha}$.

The crucial part of Macdonald polynomial definition (\ref{triangulardecomposition})-(\ref{orthogonality}) is the Macdonald density. It allowed us to systematically connect Macdonald polynomials with the corresponding Koornwinder polynomials and use them to study Macdonald polynomials.

Macdonald density is calculated as a product over all roots of the root system $R$
 \begin{equation}
     \Delta(v) := \prod_{\alpha \in R} \frac{\left(t_{2\alpha}^{1/2}e^{\alpha}(v);q_{\alpha}\right)_{\infty}}{\left(t_{\alpha}t_{2\alpha}^{1/2}e^{\alpha}(v);q_{\alpha}\right)_{\infty}}, \quad\quad (a;q)_{\infty} = \prod_{i=0}^{\infty} (1- aq^i).
\label{MacDens}
 \end{equation}
where $q_{\alpha} = q^{u_{\alpha}}$, $u_{\alpha}$ is defined from the condition that $\alpha/u_{\alpha}$ is a root in the root system $S$, $t_{\alpha}$ depends only on the length of the root $|\alpha|$ and we use $t$ when $(\alpha,\alpha)=2$, $t_l$ when $(\alpha,\alpha)=4$, $t_s$ when $(\alpha,\alpha)=1$ and $t_3$ when $(\alpha,\alpha)=6$. $t_{2\alpha}$ equals to 1 if there is no root $2\alpha$ in the root system $R$.

In the following sections we calculate Macdonald densities for all possible admissible pairs $(R,S)$ (\ref{admissible pairs}). We also point out the connection of the resulting density with Koornwinder density $\Delta_K$. We talk about Koornwinder polynomials and their definition in Appendix B.

Roots and positive roots of the root systems are written in orthogonal basis $\ve_i$, with  the following scalar product
\begin{equation}
    (\ve_i,\ve_j) = \delta_{ij}.
\end{equation}

\subsection*{$A_n$}
Root system $A_n$ corresponds to the algebra $sl_{n+1}$. There is only one possible admissible pair for the root system $A_n$: $(A_n,A_n)$. Roots and Weyl group $W$ of the root system $A_n$ are the following:
\begin{equation}
    R = \left\{\ve_i-\ve_j\,|\, 1\leq i \neq j \leq n+1\right\}, \quad\quad W \simeq S_{n+1} \end{equation}
 \begin{equation}
    R^{+} = \left\{\ve_i-\ve_j\,|\, 1\leq i < j \leq n+1\right\}.
    \end{equation}
and the Macdonald density is
\begin{equation}
    \Delta_{A_n}(x,q,t) =\, \prod_{1\leq i \neq j \leq n+1} \frac{\left(x_i/x_j; q\right)_{\infty}}{\left(t x_i/x_j;q\right)_{\infty}}
\end{equation}

\subsection*{$D_n$}
Root system $D_n$ corresponds to the algebra $so_{2n}$.
There is only one possible admissible pair for the root system $D_n$: $(D_n,D_n)$. Roots, positive roots and Weyl group $W$ of the root system $D_n$ are the following:
\begin{equation}
    R = \left\{\pm \ve_i\pm \ve_j| 1\leq i < j \leq n\right\}, \quad\quad W \simeq S_{n} \ltimes (\mathbb{Z}/2\mathbb{Z})^{n-1},\end{equation}
    \begin{equation}
    R^+ = \left\{ \ve_i\pm \ve_j| 1\leq i < j \leq n\right\}
    \end{equation}
and the Macdonald density is
\begin{align}
    \Delta_{D_n}(x,q,t) = \prod_{1\leq i < j \leq n} \frac{\left( x_i x_j;q\right)_{\infty}}{\left( t  x_i x_j  ;q\right)_{\infty}}
    \prod_{1\leq i < j \leq n} \frac{\left(x_i/x_j ;q\right)_{\infty}}{\left( t x_i/x_j  ;q\right)_{\infty}}
    \prod_{1\leq i < j \leq n} \frac{\left(x_j/x_i ;q\right)_{\infty}}{\left( t x_j/x_i  ;q\right)_{\infty}}
    \prod_{1\leq i < j \leq n} \frac{\left(1/(x_i x_j) ;q\right)_{\infty}}{\left( t/(x_i x_j)   ;q\right)_{\infty}}=\\
    \prod_{i<j}{(x_i^{\pm 1}x_j^{\pm 1};q)_{\infty}\over(tx_i^{\pm 1}x_j^{\pm 1};q)_{\infty}} \nn = \boxed{\Delta_K(x\,| \, 1,-1,q^{1/2},-q^{1/2} \,|\, q,t)},
\end{align}
where $\Delta_K$ is Koornwinder density, which we discuss in Appendix B.

\subsection*{$B_n$}
Root system $B_n$ corresponds to the algebra $so_{2n+1}$. There are two possible admissible pairs for the root system $B_n$: $(B_n,B_n)$ and $(B_n,C_n)$. Roots, positive roots and Weyl group $W$ of the root system $B_n$ are the following:
\begin{equation}
    R = \left\{\pm \ve_i | 1\leq i\leq n \right\} \cup \left\{\pm \ve_i\pm \ve_j| 1\leq i < j \leq n\right\}, \quad\quad W \simeq S_{n} \ltimes (\{\pm 1\})^n\end{equation}
\begin{equation}
    R^+ = \left\{\ve_i | 1\leq i\leq n \right\} \cup \left\{ \ve_i\pm \ve_j| 1\leq i < j \leq n\right\}.
    \end{equation}
\subsubsection*{$(B_n,B_n)$}
\begin{equation}
    \begin{array}{ll}
       q_{\ep} = q  &  q_{\pm \ep \pm \ep} = q  \\
        t_{\ep} = t_s &  t_{\pm \ep \pm \ep} =t
    \end{array}
\end{equation}

    \begin{align}
        \Delta_{B_n} (x,q,t,t_s) = & \Delta_{D_n}(x,q,t) \prod_{i=1}^{n}  \frac{\left( x_i;q\right)_{\infty}}{\left(t_s x_i ;q\right)_{\infty}}  \frac{\left( 1/x_i;q\right)_{\infty}}{\left(t_s/x_i ;q\right)_{\infty}} =  \Delta_{D_n}(x,q,t) \prod_{i=1}^{n}  \frac{\left( x_i^{\pm1};q\right)_{\infty}}{\left(t_s x_i ;q\right)_{\infty}} =\nn  \\ &
        \Delta_{D_n}(x,q,t) \prod_{i=1}^{n} \frac{\left( x_i^{\pm 2};q\right)_{\infty}}{\left( t_s x_i^{\pm1};q\right)_{\infty}\left( -x_i^{\pm1};q\right)_{\infty}\left( q^{1/2}x_i^{\pm1};q\right)_{\infty}\left( -q^{1/2} x_i^{\pm1};q\right)_{\infty}} = \nn \\
        & \boxed{\Delta_K (x\, | \, t_s, -1, q^{1/2}, -q^{1/2}\,| \,  q,t)}.
    \end{align}

\subsubsection*{$(B_n,B_n^{\vee})$}
\begin{equation}
    \begin{array}{ll}
       q_{\ep} = q^{1/2}  &  q_{\pm \ep \pm \ep} = q  \\
        t_{\ep} = t_s &  t_{\pm \ep \pm \ep} =t
    \end{array}
\end{equation}

\begin{align}
    \Delta_{(B_n,B_n^{\vee})}\, = \,& \Delta_{D_n}(x,q,t)\cdot \prod_{i=1}^{n} \frac{(x_i^{\pm1};q^{1/2})_{\infty}}{(t_s x_i^{\pm1};q^{1/2})_{\infty}} =\Delta_{D_n}(x,q,t)\cdot \prod_{i=1}^{n} \frac{(x_i^{\pm 2};q)_{\infty}}{(-x_i^{\pm1};q^{1/2})_{\infty}(t_s x_i^{\pm1};q)_{\infty}(q^{1/2}t_s x_i^{\pm1};q)_{\infty}} = \nn \\ &
    \Delta_{D_n}(x,q,t)\cdot \prod_{i=1}^{n} \frac{(x_i^{\pm 2};q)_{\infty}}{(-x_i^{\pm1};q)_{\infty}(-q^{1/2}x_i^{\pm1};q)_{\infty}(t_s x_i^{\pm1};q)_{\infty}(q^{1/2}t_s x_i^{\pm1};q)_{\infty}} = \nn \\
    & \boxed{\Delta_K(x \,|\,-1,-q^{1/2},t_s, q^{1/2}t_s\,|\, q,t)}.
\end{align}

\subsection*{$C_n$}
Root system $C_n$ corresponds to the algebra $sp_{2n}$. There are two possible admissible pairs for the root system $C_n$: $(C_n,C_n)$ and $(C_n,B_n)$. Roots, positive roots and Weyl group $W$ of the root system $C_n$ are the following:
\begin{equation}
    R =  \left\{\pm  \ve_i\pm \ve_j| 1\leq i < j \leq n\right\}\cup\left\{\pm 2\ve_i | 1\leq i\leq n \right\} , \quad\quad W \simeq S_{n} \ltimes (\mathbb{Z}/2\mathbb{Z})^n\end{equation}
    \begin{equation}
    R =  \left\{\ve_i\pm \ve_j| 1\leq i < j \leq n\right\}\cup\left\{2\ve_i | 1\leq i\leq n \right\}.\end{equation}

\subsubsection*{$(C_n,C_n)$}
\begin{equation}
    \begin{array}{ll}
       q_{2\ep} = q  &  q_{\pm \ep \pm \ep} = q  \\
        t_{2\ep} = t_l &  t_{\pm \ep \pm \ep} =t
    \end{array}
\end{equation}
      \begin{align}
        \Delta_{C_n} (x,q,t,t_l)\, =\, &  \Delta_{D_n}(x,q,t) \prod_{i=1}^{n}  \frac{\left( x_i^2;q\right)_{\infty}}{\left(t_l x_i^2 ;q\right)_{\infty}}  \frac{\left( 1/x_i^2;q\right)_{\infty}}{\left(t_l/x_i^2 ;q\right)_{\infty}} = \Delta_{D_n}(x,q,t) \prod_{i=1}^{n}  \frac{\left( x_i^{\pm 2};q\right)_{\infty}}{\left(t_l x_i^{\pm 2} ;q\right)_{\infty}} = \nn \\
     &   \Delta_{D_n}(x,q,t) \prod_{i=1}^{n} \frac{\left( x_i^{\pm 2};q\right)_{\infty}}{\left( t_l x_i^{\pm 2};q^2\right)_{\infty}\left( q t_l x_i^{\pm 2};q^2\right)_{\infty}} = \nn \\ &
    \Delta_{D_n}(x,q,t) \prod_{i=1}^{n}
    \frac{\left( x_i^{\pm 2};q\right)_{\infty}}{\left( t_l^{1/2} x_i^{\pm 1};q\right)_{\infty}\left( -t_l^{1/2} x_i^{\pm 1};q\right)_{\infty}\left( q^{1/2} t_l^{1/2} x_i^{\pm 1};q\right)_{\infty}\left( -q^{1/2} t_l^{1/2} x_i^{\pm 1};q\right)_{\infty}} = \nn \\
    & \boxed{ \Delta_{K}(x\,|\, t_l^{1/2},-t_l^{1/2}, q^{1/2}t_l^{1/2},-q^{1/2}t_l^{1/2}\,|\,q,t)}
    \end{align}

\subsubsection*{$(C_n,C_n^{\vee})$}
\begin{equation}
    \begin{array}{ll}
       q_{2\ep} = q^{2}  &  q_{\pm \ep \pm \ep} = q  \\
        t_{2\ep} = t_l &  t_{\pm \ep \pm \ep} =t
    \end{array}
\end{equation}

\begin{align}
    \Delta_{(C_n,C_n^{\vee})}\, = \,& \Delta_{D_n}(x,q,t)\cdot \frac{(x_i^{\pm 2};q^2)_{\infty}}{(t_l x_i^{\pm 2};q^2)_{\infty}} = \Delta_{D_n}(x,q,t)\cdot \frac{(x_i^{\pm 2};q)_{\infty}}{(q x_i^{\pm 2};q^2)_{\infty}(t_l x_i^{\pm 2};q^2)_{\infty}} = \nn \\
    & \Delta_{D_n}(x,q,t)\cdot  \frac{(x_i^{\pm 2};q)_{\infty}}{(q^{1/2} x_i^{\pm 1};q)_{\infty}(-q^{1/2} x_i^{\pm 1};q)_{\infty}(t_l^{1/2} x_i^{\pm 1};q)_{\infty}(-t_l^{1/2} x_i^{\pm 1};q)_{\infty}} = \nn \\
    & \boxed{\Delta_{K} (x \,|\, t_l^{1/2},-t_l^{1/2},q^{1/2},-q^{1/2}\,|\, q,t)}.
\end{align}

\subsection*{$BC_n$}
Root system $BC_n$ is irreducible, but non-reduced and can only be the first root system $R$ in the admissible pair $(R,S)$. Roots and positive roots of the root system $BC_n$ are the following:
\begin{equation}
    R = \left\{\pm \ve_i | 1\leq i\leq n \right\} \cup  \left\{\pm 2 \ve_i | 1\leq i\leq n \right\} \cup\left\{\pm \ve_i\pm \ve_j| 1\leq i < j \leq n\right\},
\end{equation}
\begin{equation}
    R^+ = \left\{ \ve_i | 1\leq i\leq n \right\} \cup  \left\{ 2 \ve_i | 1\leq i\leq n \right\} \cup\left\{ \ve_i\pm \ve_j| 1\leq i < j \leq n\right\}.
\end{equation}

\subsubsection*{$(BC_n,B_n)$}
\begin{equation}
    \begin{array}{lll}
        q_{\pm \ep} = q, & q_{\pm 2\ep} = q^2, & q_{\pm \ep \pm \ep} = q \\
        t_{\pm \ep} = a, & t_{\pm 2\ep} = b, & t_{\pm \ep \pm \ep} = t \\
    \end{array}
\end{equation}
\begin{align}
    \Delta_{(BC_n, B_n)}\, =\, & \Delta_{D_n}(x,q,t) \cdot \prod_{1\leq i\leq n}\frac{(b^{1/2}x_i^{\pm};q)_{\infty
    }}{(a b^{1/2}x_i^{\pm};q)_{\infty}}
    \prod_{1\leq i\leq n}\frac{(x_i^{\pm 2};q^2)_{\infty
    }}{(b x_i^{\pm 2};q^2)_{\infty}} \nn = \\
    &  \Delta_{D_n}(x,q,t) \cdot \prod_{1\leq i\leq n} \frac{(b^{1/2}x_i^{\pm};q)_{\infty
    } (x_i^{\pm 2};q)_{\infty
    }}{ (q x_i^{\pm 2};q^2)_{\infty
    } (a b^{1/2}x_i^{\pm};q)_{\infty} (b^{1/2}x_i^{\pm};q)_{\infty
    }(-b^{1/2}x_i^{\pm};q)_{\infty
    }} = \nn \\
    &  \Delta_{D_n}(x,q,t) \cdot \prod_{1\leq i\leq n}
    \frac{(x_i^{\pm 2};q)_{\infty}}{ (q^{1/2} x_i^{\pm};q)_{\infty} (-q^{1/2} x_i^{\pm};q)_{\infty} (a b^{1/2}x_i^{\pm};q)_{\infty} (-b^{1/2}x_i^{\pm};q)_{\infty
    }} = \nn \\
    & \boxed{\Delta_K (x \,|\, a b^{1/2}, -b^{1/2}, q^{1/2}, -q^{1/2}\,|\, q,t)}.
\end{align}

\subsubsection*{$(BC_n, C_n)$}

\begin{equation}
    \begin{array}{lll}
        q_{\pm \ep} = q^{1/2}, & q_{\pm 2\ep} = q, & q_{\pm \ep \pm \ep} = q \\
        t_{\pm \ep} = a, & t_{\pm 2\ep} = b, & t_{\pm \ep \pm \ep} = t \\
    \end{array}
\end{equation}

\begin{align}
    \Delta_{(BC_n, C_n)}\, =\, & \Delta_{D_n}(x,q,t) \cdot \prod_{1\leq i\leq n}\frac{(b^{1/2}x_i^{\pm};q^{1/2})_{\infty
    }}{(a b^{1/2}x_i^{\pm};q^{1/2})_{\infty}}
    \prod_{1\leq i\leq n}\frac{(x_i^{\pm 2};q)_{\infty
    }}{(b x_i^{\pm 2};q)_{\infty}} \nn = \\
     & \Delta_{D_n}(x,q,t) \cdot \prod_{1\leq i\leq n} \frac{(b^{1/2}x_i^{\pm};q^{1/2})_{\infty
    } (x_i^{\pm 2};q)_{\infty
    }}{(a b^{1/2}x_i^{\pm};q^{1/2})_{\infty}(b^{1/2}x_i^{\pm};q^{1/2})_{\infty}(-b^{1/2}x_i^{\pm};q^{1/2})_{\infty}}
    \nn =
    \\ & \Delta_{D_n}(x,q,t) \cdot \prod_{1\leq i\leq n} \frac{(x_i^{\pm 2};q)_{\infty
    }}{(a b^{1/2}x_i^{\pm};q)_{\infty}(a b^{1/2} q^{1/2}x_i^{\pm};q)_{\infty}(-b^{1/2}x_i^{\pm};q)_{\infty}(- b^{1/2}q^{1/2}x_i^{\pm};q)_{\infty}} = \nn \\
    & \boxed{\Delta_K (x\,|\, a b^{1/2},  a b^{1/2} q^{1/2}, -b^{1/2}, - b^{1/2} q^{1/2} \,|\, q,t)}.
\end{align}

\subsection*{$E_6$}
There is only one possible admissible pair for the root system $E_6$: $(E_6,E_6)$. Roots and positive roots of the root system $E_6$ are the following:
\begin{equation}
    R= \{\pm \ep_i \pm \ep_j | 1\leq i<j\leq 5\}\cup \{ \pm(\ep_8 - \ep_7 -\ep_6 + \sum_{i=1}^5 \pm \ep_i)/2 \,|\, \text{sum with odd number of }-\},
\end{equation}
\begin{equation}
    R^{+}= \{\pm \ep_i + \ep_j | 1\leq i<j\leq 5\}\cup \{ (\ep_8 - \ep_7 -\ep_6 + \sum_{i=1}^5 \pm \ep_i)/2 \,|\, \text{sum with odd number of }-\}.
\end{equation}

\begin{align}
    \Delta_{E_6} = \Delta_{D_{5}}(x,q,t) \cdot \prod_{\sigma}\frac{((x_8 x_7^{-1}x_6^{-1}\prod^{5}_{i=1}x_i^{\sigma})^{\pm1/2};q)_{\infty}}{(t_l(x_8 x_7^{-1}x_6^{-1}\prod^{5}_{i=1}x_i^{\sigma})^{\pm1/2};q)_{\infty}},
\end{align}
where $\sigma$ denotes sign choice in the product $\prod^{5}_{i=1}x_i^{\sigma}$, but only with an odd number of negative signs.

\subsection*{$E_7$}
There is only one possible admissible pair for the root system $E_7$: $(E_7,E_7)$. Roots and positive roots of the root system $E_7$ are the following:
\begin{equation}
    R= \{\pm \ep_i \pm \ep_j | 1\leq i<j\leq 6\}\cup \{\pm (\ep_7-\ep_8)\} \cup \{\pm( \ep_7 - \ep_8 + \sum_{i=1}^5 \pm \ep_i)/2 \,|\, \text{sum with odd number of }-\},
\end{equation}
\begin{equation}
    R^{+}= \{\pm \ep_i + \ep_j | 1\leq i<j\leq 6\}\cup \{ -(\ep_7-\ep_8)\} \cup \{-( \ep_7 - \ep_8 + \sum_{i=1}^5 \pm \ep_i)/2 \,|\, \text{sum with odd number of }-\}.
\end{equation}

\begin{align}
    \Delta_{E_7} = \Delta_{D_{6}}(x,q,t) \cdot \frac{((x_7x_8^{-1})^{\pm 1};q)_{\infty}}{(t(x_7x_8^{-1})^{\pm 1};q)_{\infty}}\cdot \prod_{\sigma}\frac{((x_7 x_8^{-1}\prod^{6}_{i=1}x_i^{\sigma})^{\pm 1/2};q)_{\infty}}{(t_l(x_7 x_8^{-1}\prod^{6}_{i=1} x_i^{\sigma})^{\pm 1/2};q)_{\infty}},
\end{align}
where $\sigma$ denotes sign choice in the product $\prod^{8}_{i=1}x_i^{\sigma}$, but only with an even number of negative signs.

\subsection*{$E_8$}
There is only one possible admissible pair for the root system $E_8$: $(E_8,E_8)$. Roots and positive roots of the root system $E_8$ are the following:
\begin{equation}
    R= \{\pm \ep_i \pm \ep_j | 1\leq i<j\leq 8\}\cup \{ \sum_{i=1}^{8}\pm \ep_i / 2 \,|\, \text{sum with even number of }-\},
\end{equation}
\begin{equation}
    R^+= \{\pm \ep_i + \ep_j | 1\leq i<j\leq 8\}\cup \{ \sum_{i=1}^{8}\pm \ep_i / 2 \,|\, \text{sum with $0$, $2$ or $4$ } - \}.
\end{equation}

\begin{align}
    \Delta_{E_8} = \Delta_{D_8}(x,q,t)\cdot \prod_{\sigma} \frac{((\prod_{i=1}^{8} x_i^{\sigma})^{ 1/2};q)_{\infty}}{(t_
    l(\prod_{i=1}^{8} x_i^{\sigma})^{1/2};q)_{\infty}},
\end{align}
where $\sigma$ denotes sign choice in the product $\prod^{6}_{i=1}x_i^{\sigma}$, but only with an even number of negative signs.

\subsection*{$F_4$}
There is only one possible admissible pair for the root system $F_4$: $(F_4,F_4)$. Roots and positive roots of the root system $F_4$ are the following:
\begin{equation}
    R=\{\pm \ep_i|1\leq i\leq 4 \}\cup \{\pm \ep_i\pm \ep_j|1\leq i <j \leq 4 \} \cup \{(\pm \ep_1\pm\ep_2\pm \ep_3 \pm \ep_4)/2\},
\end{equation}
\begin{equation}
    R^{+}=\{ \ep_i|1\leq i\leq 3 \}\cup\{-\ep_4\} \cup \{\ep_i\pm \ep_j|1\leq i <j \leq 3 \} \cup\{\pm \ep_i-\ep_4\,|\,1\leq i \leq 3\} \cup \{(\pm \ep_1\pm\ep_2\pm \ep_3 - \ep_4)/2\}.
\end{equation}

\begin{align}
    \Delta_{F_4} = \Delta_{D_4}(x,q,t)\cdot \prod_{i=1}^{4} \frac{(x_i^{\pm 1};q)_{\infty}}{(t_s x_i^{\pm 1};q)_{\infty}} \cdot \frac{((x_1^{\pm 1}x_2^{\pm 1}x_3^{\pm 1}x_4^{\pm 1})^{1/2};q)_{\infty}}{(t_l(x_1^{\pm 1}x_2^{\pm 1}x_3^{\pm 1}x_4^{\pm 1})^{1/2};q)_{\infty}}.
\end{align}

\subsection*{$G_2$}
There is only one possible admissible pair for the root system $G_2$: $(G_2,G_2)$. Roots and positive roots of the root system $G_2$ are the following:
\begin{equation}
    R = \{\pm(\ep_i-\ep_j)| 1\leq i<j\leq 3\} \cup \{\pm(2\ep_i-\ep_j-\ep_k)| \{i,j,k\}=\{1,2,3\} \},
\end{equation}
\begin{equation}
    R^{+} = \{\ep_1-\ep_2,\,-\ep_1+\ep_3,\,-\ep_2+\ep_3,\,-\ep_1-\ep_2+2\ep_3,\, \ep_1-2\ep_2+\ep_2, \, -2\ep_1+\ep_2+\ep_3\}.
\end{equation}

\begin{align}
    \Delta_{G_2} = \prod_{1\leq i < j \leq 3} \frac{(x_i^{\pm 1} x_j^{\mp 1};q)_{\infty}}{(t x_i^{\pm 1} x_j^{\mp 1};q)_{\infty}} \prod_{\{i,j,k\}=\{1,2,3\}} \frac{(x_i^{\pm 2}x_j^{\mp1}x_k^{\mp 1};q)_{\infty}}{(t_l x_i^{\pm 2}x_j^{\mp1}x_k^{\mp 1};q)_{\infty}}.
\end{align}

\section*{Appendix B. Koornwinder polynomials}
Koornwinder polynomials $P_{\lambda}$ depend on partition $\lambda$, $N$ variables $x_i$, four new parameters $(a,b,c,d)$ as well as on standard Macdonald parameters $q$ and $t$ :

\begin{equation}
    P_{\lambda}(x\,|\, a,b,c,d\,|\, q,t).
\end{equation}
In table \ref{KooMacCorr} we listed all types of Macdonald polynomials (we also denote them $P^{(R,S)}_{\lambda}(x\,|\, t_{\alpha}\,|\, q,t)$) that are generalized by Koornwinder polynomials. In Appendix A there are details of calculation of Macdonald densities and their connection to density of Koornwinder polynomials.

\begin{table}[H]
    \centering
    \begin{tabular}{|c|c|c|}
      \rule{0cm}{0.5cm}
        type of Macdonald polynomials & designation & parameters of corresponding Koornwinder  \\
       (colored with admissible pair $(R,S)$) & of Macdonald polynomials & polynomials $(a,b,c,d)$ \\[0.2cm]
       \hline
       \rule{0cm}{0.5cm}
       $(B_n,B_n)$ & $P_{\lambda}^{(B_n,B_n)} (x\,|\, t_s \,|\, q,t )$  & $(t_s, -1, q^{1/2}, -q^{1/2})$ \\[0.2cm]
       \hline
       \rule{0cm}{0.5cm}
       $(B_n,B_n^{\vee}) =(B_n,C_n) $ & $P_{\lambda}^{(B_n,C_n)} (x\,|\, t_s \,|\, q,t )$ & $(-1,-q^{1/2},t_s, q^{1/2}t_s)$ \\[0.2cm]
       \hline
       \rule{0cm}{0.5cm}
       $(C_n,C_n)$ & $P_{\lambda}^{(C_n,C_n)} (x\,|\, t_l \,|\, q,t )$ & $(t_l^{1/2},-t_l^{1/2}, q^{1/2}t_l^{1/2},-q^{1/2}t_l^{1/2})$ \\[0.2cm]
       \hline
        \rule{0cm}{0.5cm}
        $(C_n,C_n^{\vee}) = (C_n,B_n) $ & $P_{\lambda}^{(C_n,B_n)} (x\,|\, t_l \,|\, q,t )$ & $(t_l^{1/2},-t_l^{1/2},q^{1/2},-q^{1/2})$ \\[0.2cm]
       \hline
       \rule{0cm}{0.5cm}
       $(D_n, D_n)$  & $P_{\lambda}^{(D_n,D_n)} (x \,|\, q,t )$  & $(1,-1,q^{1/2},-q^{1/2})$ \\[0.2cm]
       \hline
       \rule{0cm}{0.5cm}
       $(BC_n,B_n)$ & $P_{\lambda}^{(BC_n,B_n)} (x\,|\, a, b \,|\, q,t )$  & $(a b^{1/2}, -b^{1/2}, q^{1/2}, -q^{1/2})$ \\[0.2cm]
       \hline
       \rule{0cm}{0.5cm}
       $(BC_n, C_n)$ & $P_{\lambda}^{(BC_n,C_n)} (x\,|\, a,b \,|\, q,t )$  & $( a b^{1/2},  a b^{1/2} q^{1/2}, -b^{1/2}, - b^{1/2} q^{1/2})$ \\[0.2cm]
       \hline
    \end{tabular}
     \caption{Correspondence between Koornwinder and Macdonald polynomials}
    \label{KooMacCorr}
\end{table}

The Koornwinder polynomials are defined from the triangular decomposition and from the orthogonality w.r.t. the scalar product given by the Koornwinder density
\be
\Delta_K:=\prod_{i=1}^n{(x_i^{\pm 2};q)\over (ax_i^{\pm 1},bx_i^{\pm 1},cx_i^{\pm 1},dx_i^{\pm 1};q)}
\prod_{i<j}{(x_i^{\pm 1}x_j^{\pm 1};q)\over(tx_i^{\pm 1}x_j^{\pm 1};q)}
\ee
With this density, the scalar product is given by
\be
\Big<A(x)\Big|B(x)\Big>={(q;q)^n\over 2^nn!} \left[A(x)B(x^{-1})\Delta\right]_0
\ee
and the Koornwinder polynomials are defined from the expansion
\be
P_\lambda=\xi_\lambda+\sum_{\mu<\lambda}K_{\lambda\mu}\xi_{\mu}
\ee
where the condition $\mu<\lambda$ implies the dominant ordering.

In case of root system $C_n$: $\lambda\geq \mu$ $\Longleftrightarrow$ $|\lambda|\,\equiv\, |\mu|\,(\text{mod }2)$ and $\sum_{k=1}^{i}\lambda_ k\geq \sum_{k=1}^{i}\mu_k$ for all $i=1,2, \dots$. Partitions of even numbers $|\lambda| = 2n$ are dominant to the empty partition $\emptyset$: $\lambda>\emptyset$.

\paragraph{Koornwinder Hamiltonian.}
Koornwinder polynomials are eigenfunctions
\be
\hat{\cal D}\cdot P_\lambda(x\,|\, a,b,c,d\,|\, q,t)=
\Lambda_\lambda P_\lambda(x\,|\, a,b,c,d\,|\, q,t)
\ee
of the difference operator $\hat{\cal D}$:
\be
\hat{\cal D}:=\sum_{i=1}^nt{(1-ax_i)(1-bx_i)(1-cx_i)(1-dx_i)\over \alpha\xi(1-x_i^2)(1-qx_i^2)}\prod_{j\ne i}
{(1-tx_ix_j)(1-tx_i/x_j)\over (1-x_ix_j)(1-x_i/x_j)}\left(q^{x_i\p_i}-1\right)+\nn\\
+\sum_{i=1}^nt{(1-ax_i^{-1})(1-bx_i^{-1})(1-cx_i^{-1})(1-dx_i^{-1})\over \alpha\xi(1-x_i^{-2})(1-qx_i^{-2})}\prod_{j\ne i}
{(1-tx_i^{-1}x_j^{-1})(1-tx_j/x_i)\over (1-x_i^{-1}x_j^{-1})(1-x_j/x_i)}\left(q^{-x_i\p_i}-1\right),\nn\\
\alpha:=\sqrt{abcd\over q},\,\, \xi = t^n
\ee
with the eigenvalues
\be
\Lambda_\lambda=\sum_{j=1}^n\{\alpha t^{n-j}q^{\lambda_j/2}\}\{q^{\lambda_j/2}\}.
\ee

\noindent \textbf{Koornwinder polynomials for antisymmetric partitions.}

\noindent
It is very convenient to use explicit expressions for Koornwinder polynomials to get Macdonald polynomials.

For example one of the wonderful formulas was provided in \cite{Sh2}

\begin{equation}
  \boxed{  P_{(1^r)}(x \mid a, b, c, d \mid q, t) =
\sum_{k,l,i,j \geq 0} (-1)^{i+j} E_{r-2k-2l-i-j}(x)
\widehat{c}_e'(k, l; t^{n-r+1+i+j})
\widehat{c}_o(i, j; t^{n-r+1}),}
\end{equation}
where
\begin{equation}
    \widehat{c}_e'(k, l; s) =
\frac{(tc^2/\alpha^2; t^2)_k (sc^2t; t^2)_k (s^2c^4/t^2; t^2)_k}
{(t^2; t^2)_k (sc^2/t; t^2)_k (s^2a^2c^2/t; t^2)_k}
\frac{(1/c^2; t)_l (s/t; t)_{2k+l}}{(t; t)_l (sc^2; t)_{2k+l}}
\frac{1 - st^{2k+2l-1}}{1 - st^{-1}} a^{2k} c^{2l}.
\end{equation}
and
\begin{equation}
    \widehat{c}_o(i,j; s) =
\frac{(-a/b; t)_i (scd/t; t)_i}{(t; t)_i (-sac/t; t)_i}
\frac{(s; t)_{i+j} (-sac/t; t)_{i+j} (s^2 a^2 c^2/t^3; t)_{i+j}}
{(s^2 abcd/t^2; t)_{i+j} (sac/t^{3/2}; t)_{i+j} (-sac/t^{3/2}; t)_{i+j}} \frac{(-c/d; t)_j (sab/t; t)_j}{(t; t)_j (-sac/t; t)_j} b^i d^j.
\end{equation}
and $E_r(x)$ are defined with the generating function
\begin{equation}
    E(x \mid y) = \prod_{i=1}^{n} (1 - yx_i)(1 - y/x_i) = \sum_{r \geq 0} (-1)^r E_r(x) y^r .
\end{equation}
$E_r(x)$ are just Schur functions $S_{[1^r]}(p_m)$ expressed with power sum symmetric functions $p_m$, where one uses $p_m = \sum_{i=1}^{n} x_i^m+x_i^{-m}$. Some formulas for Macdonald and Koornwinder polynomials can also be found in \cite{Sh1,Sh2,Ok}.

\section*{Appendix C. Macdonald polynomials at the point $q^{2\rho_k}$}
In this section we return to the notation of this paper where we use squares of Macdonald parameters.

We list some examples how Macdonald polynomials do not factorize at the point $q^{2\rho_k}$, which is a refined version of the point $q^{2\rho}$ in the definition of quantum dimensions (\ref{qd}).

\subsection*{Macdonald polynomials for root system $C_n$ at the point $q^{2\rho_k}$}
Values of the symplectic Macdonald polynomials at the point $q^{2\rho_k}$ and, to compare, the limit of them to the symplectic characters. Here we use our standard notation $\{x\} = x-x^{-1}$, $[x]_t = \frac{t^n-t^{-n}}{t-t^{-1}}$, $ \xi_n = t_l t^{n-1}$.

Power sum symmetric functions $p_m (x_1,\dots, x_n) := \sum_{i=1}^n x_i^m+x_i^{-m}$ at the point $x = q^{2\rho_k(C_n)} = t_l^2 t^{2(i-1)}$ are the following
\begin{equation}
    p_m \left(x = q^{2\rho_k(C_n)} \right) = \frac{[m\,n]_t}{[m]_t} \frac{\{t_l^{4m}t^{2m(n-1)}\}}{\{t_l^{2m}t^{m(n-1)}\}}.
\end{equation}
And the definition
\begin{equation}
    p_{\lambda} = \prod_{i=1}^{l(\lambda)} p_{\lambda_i}
\end{equation}
works for power sum symmetric functions at a special point too.

\begin{equation}
   P_{\lambda}^{C_n}\left(x = q^{2\rho_k}\,|\,t_l^2 \,|\, q^2, t^2\right)  \,\, \xrightarrow[]{t\,\rightarrow\, q,\, t_l \rightarrow\, q} \,\, Sp_{\lambda}(q^{2\rho}),
\end{equation}
where $Sp_{\lambda}$ are characters of $sp_{2n}$ (symplectic) algebra. We call them symplectic Schur symmetric functions and discuss them in the section \ref{SchursSection}.

\begin{align}
   P_{[1]}^{C_n}(q^{\rho_k}) & = [n]_t  \frac{\{t_l^4\,t^{2n-2}\}}{\{t_l^2\,t^{n-1}\}},
       \\
   Sp_{[1]}(q^\rho) & = \frac{[n]_q[2n+2]_q}{[n+1]_q},  \\
 P_{[2]}^{C_n}(q^{\rho_k}) & = [n]_t{\{q\,t^n\}\over\{q\,t\}}\left({\{t_l^3\xi_n^3\} \over \{t_l\xi_n\}}+{\{q^{-1} \xi_n\}\over \{q\xi_n\}}\right),
     \\
   Sp_{[2]}(q^\rho) & = \frac{[n]_q[2n+1]_q[2n+4]_q}{[2]_q[n+2]_q} =  \overline{D}^{C_n}_{\Adj} ,   \\
   P_{[1,1]}^{C_n}(q^{\rho_k}) & =    \frac{[n]_t}{[2]_t} \frac{\{q\,t\}\{t^{n-1}\} \{q^2 \, t_l^2 \, t^{2n-4}\}}{\{q\,t_l\,t^{n-1}\}\{q\,t_l\,t^{n-2}\}\{q\,t_l^2\,t^{2n-3}\}} +
   \frac{1}{2} [n]_t^2 \frac{\{t_l^4 t^{2n-2}\}^2}{\{t_l^2 t^{n-1}\}^2}
   - \frac{1}{2} \frac{[2n]_t}{[2]_t} \frac{\{ t_l^8\,t^{4n-4}\}}{\{ t_l^4\,t^{2n-2}\}},  \\
Sp_{[1,1]}(q^\rho) & = \frac{[n-1]_q[2n+1]_q[2n+2]_q}{[2]_q[n+1]_q}, 
\end{align}
\begin{align}
    P_{[1,1,1]}^{C_n} (q^{\rho_k}) & = [n]_t \frac{\{t_l^4 t^{2n-2}\}}{\{t_l^2 t^{n-1}\}}  \left( \frac{[n-2]_t\{t^{n-1}\}\{q\,t\}\{q^2 t_l^2 t^{2(n-3)}\}}{[2]_t\{q\,t_l\,t^{n-2}\}\{q\,t_l^2 t^{2n-5}\}\{q\,t_lt^{n-3}\}} -[n-1]_t \frac{\{q\,t_l\}}{\{q\,t_l\,t^{n-2}\}} \right) + \\
\nn & \quad + \frac{1}{3}\,\frac{[3n]_t}{[3]_t} \frac{\{ t_l^{12}\,t^{6(n-1)}\}}{\{ t_l^6\,t^{3(n-1)}\}}-\frac{1}{2}\,\frac{[2n]_t\,[n]_t}{[2]_t} \frac{\{ t_l^8\,t^{4(n-1)}\}}{\{ t_l^4\,t^{2(n-1)}\}} \frac{\{t_l^4 t^{2(n-1)}\}}{\{t_l^2 t^{n-1}\}} +\frac{1}{6}\,[n]_t^3 \frac{\{t_l^4 t^{2(n-1)}\}^3}{\{t_l^2 t^{n-1}\}^3},  \\
  Sp_{[1,1,1]}(q^\rho) & = \frac{[n-2]_q[2n]_q[2n+1]_q[2n+2]_q}{[2]_q[3]_q[n+1]_q} 
\end{align}

\subsection*{Macdonald polynomials for root system $B_n$ at the point $q^{\rho_k}$}
Power sum symmetric functions $p_m (x_1,\dots, x_n) := \sum_{i=1}^n x_i^m+x_i^{-m}$ at the point $x = q^{2\rho_k(B_n)} = t^{2(i-1)}t_s^2$ are the following
\begin{equation}
    p_m \left(x = q^{2\rho_k(B_n)} \right) = \frac{[m\,n]_t}{[m]_t} \frac{\{t_s^{2m}t^{2m(n-1)}\}}{\{t_s^{m}t^{m(n-1)}\}}.
\end{equation}
The limit of $B_n$-colored Macdonald polynomials at a point $q^{2\rho_k}$ when the parameters $t$ and $t_s$ go to $q$ are the corresponding quantum dimensions: characters of $so_{2n+1}$ Lie algebra --- orthogonal Schur functions $So_{\lambda}(x)$ at the point $x = q^{2\rho}$:

\begin{equation}
   P_{\lambda}^{B_n}\left(x = q^{2\rho_k}\,|\,t_s^2\,|\,q^2,t^2\right)  \,\, \xrightarrow[]{t\,\rightarrow\, q,\, t_s \rightarrow\, q} \,\, So_{\lambda}(q^{2\rho}).
\end{equation}
We listed some examples of these Schur functions in the section 
\ref{SchursSection}.

\begin{align}
      & P_{[1,1]}^{B_n}(q^{2\rho_k})  =  \,[n]_t[n-1]_t \frac{\{t_s^{2}\,t^{2(n-1)}\}}{\{t_s \,t^{n-1}\}} \frac{ \{t_s\}  \{q^2 t^{2 (n-2)}\}}{\{q\, t^{n-2}\}\{q \,t_s t^{2 n-4}\}} - \frac{[n]_t \{q\}}{\{q\,t^{n-1}\}} - \frac{[n]_t}{[2]_t} \frac{\{t^{n-1}\}\{q^2\,t^{2n-4}\}\{t_s^2\,q^{-1}t^{-1}\}}{\{qt^{n-1}\}\{qt^{n-2}\}\{qt_s^2\,t^{2n-3}\}}+ \nn\\
     & + \frac{[n]_t[n-1]_t}{[2]_t} \frac{\{t_s\}\{t\,t_s\}\{q\,t_s^2 t^{2n-4}\}\{q^2\,t^{2n-2}\}\{q^2\,t^{2n-4}\}}{\{q\,t^{n-1}\}\{q\,t^{n-2}\}\{q\,t_s\,t^{2n-3}\}\{q\,t_s\,t^{2n-4}\}\{q\,t_s^2\,t^{2n-3}\}} +\frac{1}{2}\left([n]_t \frac{\{t_s^2 t^{2(n-1)}\}}{\{t_s t^{n-1}\}}\right)^2 -\frac{1}{2} \frac{[2n]_t}{[2]_t} \frac{\{t_s^4 t^{4(n-1)}\}}{\{t_s^2 t^{2(n-1)}\}}\\
     & So^{B_n}_{[1,1]}(q^{2\rho}) =  \frac{[n+1/2]_q [2n]_q [2n-3]_q}{[n-3/2]_q[2]_q}  =  D_{\Adj}^{B_n} .
\end{align}

\end{document}